\pdfoutput=1

\documentclass[journal, 10pt]{IEEEtran}
\usepackage{booktabs}
\usepackage{amssymb}
\usepackage{color, cite}
\usepackage{algorithm}
\usepackage{mathrsfs}
\usepackage{graphicx}
\usepackage{amsmath,bm,amssymb}
\usepackage{multicol}
\usepackage{CJK}
\usepackage{indentfirst}
\usepackage{stfloats}
\usepackage{bm}
\usepackage{amsmath,bm}
\usepackage{amsmath}

\usepackage{cuted}
\usepackage{float}
\usepackage{multirow}
\usepackage{changepage}
\usepackage{amsfonts}
\usepackage{stfloats}
\usepackage{array}
\usepackage{titletoc}
\usepackage{algorithm}
\usepackage{algorithmic}
\usepackage{stfloats}
\usepackage{color}

\begin{document}

\title{ \huge{Large-Scale RIS Enabled Air-Ground Channels: Near-Field Modeling and Analysis} }

\author{Hao Jiang, \IEEEmembership{Member, IEEE}, Wangqi Shi, Zaichen Zhang, \IEEEmembership{Senior Member, IEEE}, \\Cunhua Pan, \IEEEmembership{Senior Member, IEEE}, Qingqing Wu, \IEEEmembership{Senior Member, IEEE}, Feng Shu, \IEEEmembership{Member, IEEE}, Ruiqi Liu, \\and Jiangzhou Wang, \IEEEmembership{Fellow, IEEE}

\thanks{This work is supported by the Fundamental Research Funds for the Central Universities (No. 2242022k60001), the NSFC projects (No. 61960206005 and 62101275), and the National Natural Science Foundation of China (Nos. U22A2002). }
\thanks{H. Jiang and W. Shi are with the School of Artificial Intelligence/School of Future Technology, Nanjing University of Information Science and Technology, Nanjing 210044, P. R. China. H. Jiang is also with the National Mobile Communications Research Laboratory, Southeast University, Nanjing 210096, P. R. China (e-mails: \{jianghao, shiwangqi\}@nuist.edu.cn). }
\thanks{Z. Zhang and C. Pan are with the National Mobile Communications Research Laboratory, Southeast University, Nanjing 210096, China. Z. Zhang is also with the Purple Mountain Laboratories, Nanjing 211111, China (e-mails: \{zczhang, cpan\}@seu.edu.cn). }
\thanks{Q. Wu is with the Department of Electronic Engineering, Shanghai Jiao Tong University, 200240, China (e-mail: qingqingwu@sjtu.edu.cn).}
\thanks{F. Shu is with the School of Information and Communication Engineering, Hainan University, Haikou 570228, China; and also with School of Electronic and Optical Engineering, Nanjing University of Science and Technology, Nanjing 210094, China (email: shufeng0101@163.com). }
\thanks{R. Liu is with the wireless and computing research institute, ZTE Corporation, beijing 100029, China (e-mail: richie.leo@qq.com).}
\thanks{J. Wang is with the School of Engineering, University of Kent, CT2 7NT Canterbury, U.K. (e-mail: j.z.wang@kent.ac.uk).}
}

\maketitle

\begin{abstract}

Existing works mainly rely on the far-field planar-wave-based channel model to assess the performance of reconfigurable intelligent surface (RIS)-enabled wireless communication systems. However, when the transmitter and receiver are in near-field ranges, this will result in relatively low computing accuracy. To tackle this challenge, we initially develop an analytical framework for sub-array partitioning. This framework divides the large-scale RIS array into multiple sub-arrays, effectively reducing modeling complexity while maintaining acceptable accuracy. Then, we develop a beam domain channel model based on the proposed sub-array partition framework for large-scale RIS-enabled UAV-to-vehicle communication systems, which can be used to efficiently capture the sparse features in RIS-enabled UAV-to-vehicle channels in both near-field and far-field ranges. Furthermore, some important propagation characteristics of the proposed channel model, including the spatial cross-correlation functions (CCFs), temporal auto-correlation functions (ACFs), frequency correlation functions (CFs), and channel capacities with respect to the different physical features of the RIS and non-stationary properties of the channel model are derived and analyzed. Finally, simulation results are provided to demonstrate that the proposed framework is helpful to achieve a good tradeoff between model complexity and accuracy for investigating the channel propagation characteristics, and therefore providing highly-efficient communications in RIS-enabled UAV-to-vehicle wireless networks.

\end{abstract}

\IEEEpeerreviewmaketitle

\section{Introduction}

\subsection{Background}

Over the past few years, air-to-ground communications have gradually been envisioned as one of the indispensable parts for the sixth generation (6G) mobile networks \cite{Saad,JHNetworks}. To effectively enhance the performance of air-to-ground wireless communication systems, reconfigurable intelligent surface (RIS), which is a two-dimensional (2D) electromagnetic material panel consisting of square metallic patches, can be explored to manipulate the impinging waves in a programmable manner for altering the channel realizations by converting uncontrollable signal propagation environments into partially controllable ones \cite{QQWu,QQWu1}. Within this perspective, the integration of RIS technology and unmanned aerial vehicle (UAV)-to-vehicle communications has become a consensus for 6G wireless communication systems.

\subsection{Related Works}

As the basic fundamental to constructing a RIS-enabled UAV-to-vehicle communication system, the characterization of RIS-enabled channels that takes into account the underlying propagation characteristics is urgently needed \cite{Yildirim}. As a result of the substantial growth in the size of the RIS array, the dimension of passive elements in systems utilizing RIS will undergo a significant enlargement. This means that RIS-enabled wireless channels are more likely to operate in near-field ranges, where channel modeling based on the planar wavefront is no longer satisfied \cite{ss}. This phenomenon will lead to high hardware costs and increased computational complexity for evaluating the system performance, and therefore the RIS-enabled channel matrix will be extremely complicated. According to \cite{cwhuang}, the RIS-enabled wireless channels exhibit sparsity natures in beam domains. This highlights the importance of understanding the properties of beam domain channels and developing realistic beam domain channel models (BDCMs) for RIS-enabled communication systems. Specifically, BDCM is defined to make use of the sparsity characteristics of RIS-enabled channels to reduce modeling complexity, which was proposed by transforming the channel from the antenna domain into the beam domain with the assistance of beamforming matrices \cite{BasarPoor}. This approach is inherently suitable for RIS-enabled scenarios as the correlation of channel elements in BDCM continually decreases with increasing RIS unit dimensions. Up to now, numerous studies have focused on beam domain channel modeling and analysis of propagation characteristics. However, in reality, the existing beam domain channel models, such as those proposed in \cite{FLai} and \cite{HChang}, mainly regard RIS reflecting units as one point, which neglect the deviations of phases of signals arriving at different RIS units, especially in near-field range \cite{Alexandropoulos,Yjiang}, thereby demonstrating that they are not suitable for accurately evaluating the RIS-enabled wireless communication system performance. Therefore, a realistic beam domain channel model for RIS-enabled UAV-to-vehicle communications considering the near-field effect with low computational complexity and acceptable accuracy, which is suitable for information theory and signal processing researches of RIS-enabled UAV-to-vehicle systems, should be an open research direction \cite{hZhang}.

Owing to the large size of RIS array aperture, communication scenarios can be classified as far-field range and near-field ranges. The classification criteria is whether the distance between the transmitting/receiving antenna array and RIS is larger than the Fraunhofer distance, which is directly proportional to the square of the antenna aperture and inversely proportional to the signal wavelength \cite{MCui2023}. Existing researches have shown that the investigation of RIS-enabled channel propagation statistics based on the near-field spherical-wave model has obvious differences from that based on the far-field planar-wave-based model \cite{Pan1,Pan2}. Therefore, the Fraunhofer distance criterion must be derived in advance to determine the boundary of the far-field and near-field ranges for measuring the channel propagation statistics \cite{YLiu}. Specifically, when the distance from a transmit/receive antenna array to RIS array exceeds the Fraunhofer distance in RIS-enabled wireless channels, which corresponds to the far-field range, the planar-wave-based model needs to be considered to characterize its propagation mechanism. In this context, the computing complexity is low because the phase differences among different RIS units do not need to be distinguished. However, if the number of transmit/receive antenna elements or RIS units is sufficiently large, resulting in a distance between the transmit/receive antenna array and the RIS array that is shorter than the Fraunhofer distance, the transmitter and receiver will be in the near-field range. In this context, the planar wavefront is not suitable to describe the propagation mechanism owing to the limited accuracy. Instead, the spherical-wave-based model should be considered \cite{ZhangY}.

In RIS-enabled wireless networks, owing to the manipulation of a RIS array on the impinged waves, the propagation mechanism of the subchannel between a transmitter and RIS will be different from that of the subchannel between the RIS and receiver. This implies that conventional channel modeling methods without considering a RIS is not suitable for modeling the RIS-enabled wireless channels \cite{EBasar}. Current works have proposed several channel models to assess the RIS-enabled communication system performance, which can be mainly classified into the RIS-enabled cascaded channel models and RIS-enabled spatial scattering channel models \cite{zczhang}. The former ones mean that the channel between a transmitter and a receiver is divided into a transmitter-RIS subchannel and a RIS-receiver subchannel \cite{JHTWC2021}. Although this kind of channel modeling method is able to capture the underlying physical features of each subchannel flexibly, in reality, the reconfigurable statistics of the RIS units are related to many physical features of RIS, which cannot be simplified to a constant. Furthermore, the transmitter-RIS subchannel and RIS-receiver subchannel are correlated, while the correlation is ignored in this modeling method. It is worth noting that in order to compensate for the high path loss, RIS necessitates a significant number of units. However, augmenting the number of RIS units, particularly in large-scale RIS deployments, results in larger array dimensions and subsequently an increased Fraunhofer distance for the RIS array. Therefore, an optional solution for achieving this goal is to adopt the sub-array partition-based modeling paradigm \cite{Najafi}. \cite{JTwc} introduced a sub-array pattern framework based on the RIS-enabled cascaded channel modeling method, in reality, we acknowledged that the physical features of the transmitter-RIS subchannel and RIS-receiver subchannel have great impact on the propagation statistics of RIS-enabled wireless channels, which were not considered in that study. This poses challenges in accurately capturing the performance of RIS-enabled wireless communication systems. On the other hand, the RIS-enabled spatial scattering channel modeling method assumes that the diagonal distance of the RIS are much larger than the wavelength, treating each unit as a reflector in RIS-enabled propagation environment \cite{Xiong2022}. This kind of channel modeling method does avoid the limitations of the RIS-enabled cascaded channel modeling method, however, there is still a lack of related work on the design of the framework of the sub-array pattern scheme by utilizing the RIS-enabled scattering channel modeling method. This gap in knowledge poses challenges in achieving high-efficiency communications in RIS-enabled wireless channels.

\subsection{Main Contributions}

To address the above issues, we present a novel analytical framework for sub-array partitioning and subsequently develop a beam domain channel model specifically for near-field communication scenarios in large-scale RIS-enabled UAV-to-vehicle communications. The main contributions are summarized as follows:

\begin{itemize}
\item  Our proposed framework involves the partitioning of the large-scale RIS array into multiple sub-arrays, which ensures that each sub-array is satisfied by the planar wavefront. This approach results in accurate and low-complexity analysis of channel propagation statistics across spatial, time, and frequency domains. We validate the effectiveness of this framework in achieving accurate and efficient investigations of channel propagation characteristics.
\end{itemize}

\begin{itemize}
\item  We conduct a comparison between the modeling performances of RIS-enabled channels using our provided sub-array partition framework and the traditional methods, specifically planar wavefront. This comparison validates the superior advantages of our proposed solution in terms of high accuracy and acceptable complexity. These findings provide theoretical support for the design of RIS-enabled UAV-to-vehicle communication systems.
\end{itemize}

\begin{itemize}
\item  We develop a beam-domain channel model for large-scale RIS-enabled UAV-to-vehicle communication environments using the sub-array partition framework. This model is derived from a transformation of the geometry-based channel model using a specially designed block transformation matrix. As a result, it effectively captures the sparsity characteristics in RIS-enabled UAV-to-vehicle channels. This enables highly-efficient UAV-to-vehicle communication within the near-field range.
\end{itemize}

\begin{itemize}
\item  We explore the statistical properties of the proposed beam domain RIS-enabled UAV-to-vehicle channel model utilizing the sub-array partition framework. This investigation includes analyzing the spatial cross-correlation functions (CCFs), temporal auto-correlation functions (ACFs), frequency correlation functions (CFs), and channel capacity. The impact of physical features of RIS and non-stationary properties on the channel propagation characteristics are derived and studied. These comprehensive analysis provides valuable insights into the behavior of RIS-enabled UAV-to-vehicle channels.
\end{itemize}

\begin{figure}[!t]
\centering
\includegraphics[width=8.5cm]{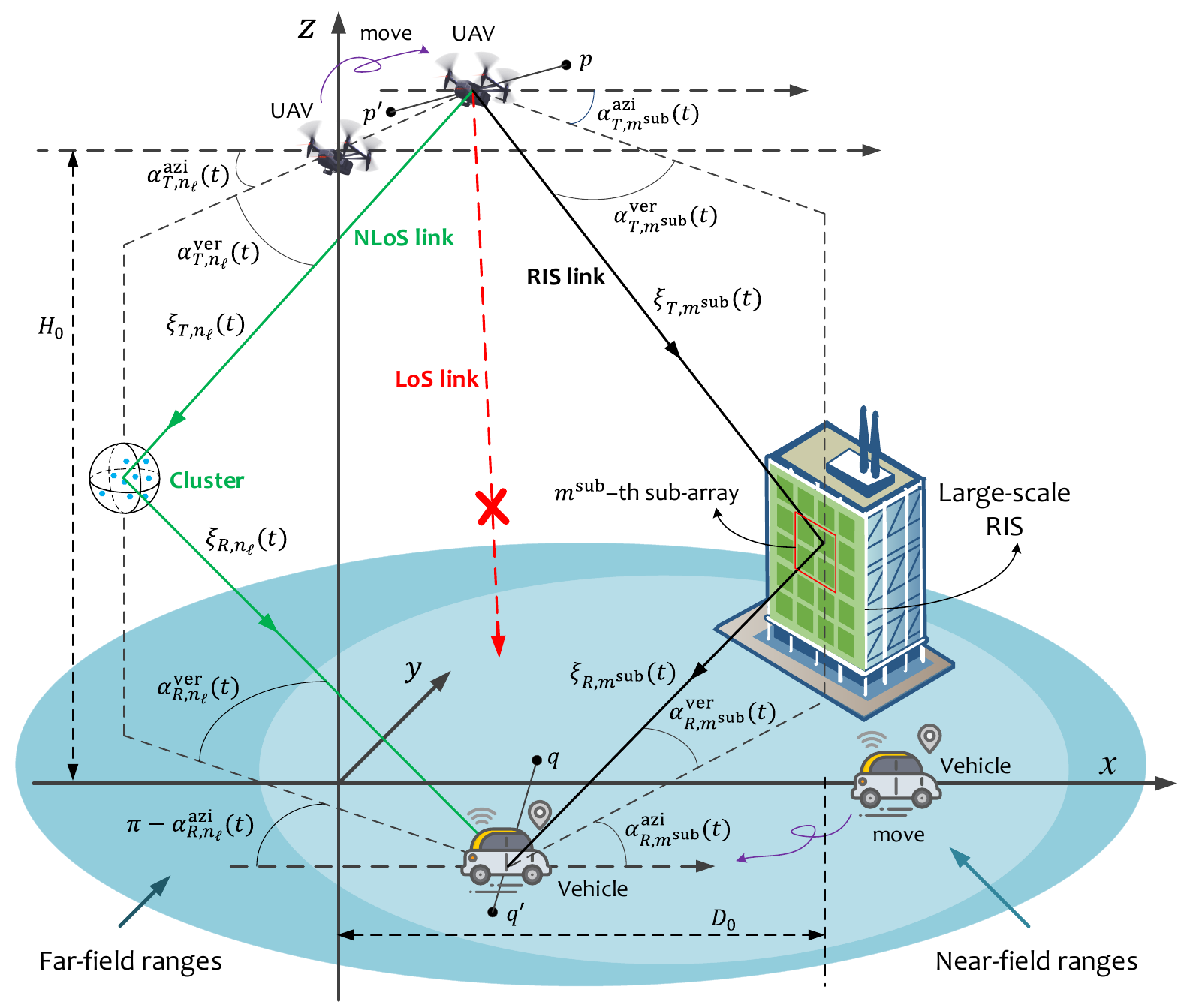}
\caption{Illustration of the propagation mechanism in the proposed RIS-enabled UAV-to-vehicle communication scenarios.}
\end{figure}

The organization of the rest of the paper is as follows. Section II presents the sub-array partition analytical framework proposed in this paper. Section III focuses on the channel modeling for RIS-enabled UAV-to-vehicle communication scenarios, including the derivation and analysis of the complex CIRs for RIS and NLoS components. In Section IV, we explore the propagation statistics of the channel model by utilizing the sub-array pattern framework. Section V presents the simulation results and discussions. Finally, Section VI draws conclusions.


\section{Proposed Sub-Array Partition Scheme}

\begin{table}[!t]
      \footnotesize  
      \centering
      \caption{Explications of Crucial Model Parameters in RIS-enabled UAV-to-Vehicle Channels}
      \begin{tabular}{ |m{1.15cm}<{\centering}|m{6.3cm}<{\centering}| }
      \hline
        $P$, $Q$  &  Numbers of ULAs at UAV and vehicle sides  \\
      \hline
        $\delta_T$, $\delta_R$  &  Spacings between two neighboring antennas in UAV/vehicle ULAs   \\
      \hline
        $\psi^\text{azi}_T$, $\psi^\text{ver}_T$  &  Azimuth and vertical angles of UAV antenna array    \\
      \hline
        $\psi^\text{azi}_R$, $\psi^\text{ver}_R$  &  Azimuth and vertical angles of vehicle antenna array  \\
       \hline
        $t$  &  Motion time of UAV and vehicle   \\
      \hline
        $v_T$, $v_R$  &  Motion speeds of UAV and vehicle    \\
      \hline
        $\eta^\text{azi}_T$, $\eta^\text{azi}_R$  &  Motion directions of UAV and vehicle relative to the positive direction of the $x$-axis    \\
      \hline
        $\eta^\text{ver}_T$  &  Motion directions of UAV along the vertical plane    \\
     \hline
        $D_0$  &  Distance from projection of midpoint of UAV ULA in the azimuth plane to the midpoint of vehicle ULA   \\
      \hline
        $H_0$  &  Distance from midpoint of UAV ULA to azimuth plane   \\
      \hline
        $M_x$, $M_z$  &  Number of RIS reflection elements along the horizontal and vertical dimensions    \\
      \hline
        $d_M$   &  Spacings between two neighboring elements in RIS array    \\
      \hline
      \end{tabular}
\end{table}

As depicted in Fig. 1, we consider a RIS-enabled UAV-to-vehicle wireless communication system which suffers from severe performance deterioration due to the blocking line-of-sight (LoS) path between a mobile UAV and a moving vehicle. To mitigate the increased path loss, a large-scale RIS is employed on the facade of a building. Here, we assume that the dimension of the RIS is sufficiently large, which ensures that both the UAV-RIS subchannel and RIS-vehicle subchannel are within the near-field range. The explication of the crucial model parameters are collected in Table I. Here, we denote the distance vectors from the $p$-th ($p = 1, 2, \cdots, P$) element of the UAV uniform linear array (ULA) and the $q$-th ($q = 1, 2, \cdots, Q$) element of the vehicle ULA to the origin of the global coordinate system as follows:
\allowdisplaybreaks[4]
\begin{eqnarray}
\textbf{d}_{T,p} \hspace*{-0.225cm}&=&\hspace*{-0.225cm} \frac{P-2p+1}{2} \delta_T \, \left[ \begin{array}{ccc} \, \cos\psi^\text{ver}_T \cos\psi^\text{azi}_T \, \\ \cos\psi^\text{ver}_T \sin\psi^\text{azi}_T \\  \sin\psi^\text{ver}_T  \end{array} \right]   \, , \\ [0.125cm]
\textbf{d}_{R,q} \hspace*{-0.225cm}&=&\hspace*{-0.225cm} \frac{Q-2q+1}{2} \delta_R \, \left[ \begin{array}{ccc} \, \cos\psi^\text{ver}_R \cos\psi^\text{azi}_R \, \\ \cos\psi^\text{ver}_R \sin\psi^\text{azi}_R \\ \sin\psi^\text{ver}_R  \end{array} \right]   \, .
\end{eqnarray}
Considering the abundant scattering propagation containing $N$ clusters, we only depict the $n$-th ($n=1,2,...,N$) cluster in Fig. 1 for clarity. Typically, current channel models for RIS-enabled systems utilize the geometry-based approach to determine distance and angle parameters for all RIS units. While this method ensures accurate modeling, it also leads to significant computational complexity due to the extensive number of RIS units needed for path loss compensation \cite{QYang}. In certain instances, the planar-wave-based model has been employed to reduce channel modeling in RIS-enabled systems operating in the far-field range. However, practical scenarios often do not meet the far-field range criteria, particularly in large-scale RIS-enabled systems. This is due to the fact that RIS size becoming comparable to the propagation distance, rendering it non-negligible, and the terminals potentially moving within the near-field range of the RIS. To address this issue, as depicted in Fig. 2, we provide a sub-array partition analytical framework that divides the large-scale RIS into multiple sub-arrays. Each sub-array, with a smaller array aperture, is suitable for the planar-wave-based model. As a result, once the distance and angle parameters of one RIS unit are determined, the parameters of the remaining units can be inferred correspondingly, effectively minimizing the complexity of the channel modeling process. Our sub-array partition scheme adopts the Fraunhofer distances as theoretical boundaries between the far-field and near-field ranges to perform the partition.

\begin{figure}[!t]
\centering
\includegraphics[width=6.5cm]{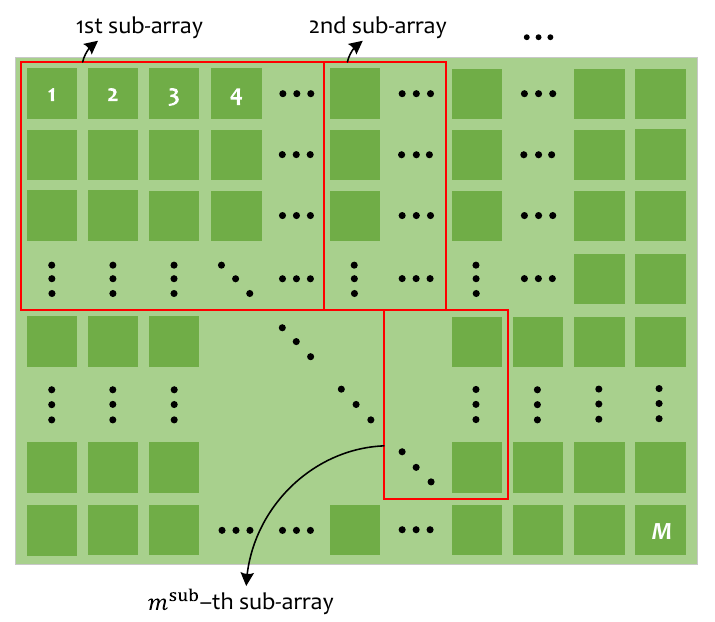}
\caption{Diagram of the proposed sub-array partition analytical framework.}
\end{figure}

As both the UAV transmitter and vehicle receiver are in motion, the proposed channel model demonstrates time-varying characteristics in the time domain. \cite{Zcui1,Zcui2}. Consequently, the distance vectors from the origin of the global coordinate system to the midpoints of the UAV ULA and vehicle ULA are correspondingly expressed as $\textbf{d}_T(t) = [ \, d_{T,x} (t), d_{T,y} (t), d_{T,z} (t) \, ]^\text{T}$ and $\textbf{d}_R(t) = [ \, d_{R,x} (t), d_{R,y} (t), 0 \, ]^\text{T}$, in which $d_{T,x} (t) = v_T t \cos\eta^\text{ver}_T\cos\eta^\text{azi}_T$, $d_{T,y} (t) = v_T t \cos\eta^\text{ver}_T\sin\eta^\text{azi}_T$, $d_{T,z} (t) = H_0 + v_T t \sin\eta^\text{ver}_T$, $d_{R,x} (t) = D_0 + v_R t \cos\eta^\text{azi}_R$, and $d_{R,y} (t) = v_R t \sin\eta^\text{azi}_R$. Furthermore, the distance vector from the origin of coordinate to the midpoint of the RIS array is denoted by $\textbf{d}_\text{RIS} = [ \, x_\text{RIS}, y_\text{RIS}, z_\text{RIS} \, ]^\text{T}$. Based on the above expressions, we can respectively denote the propagation distances from the midpoints of the UAV and vehicle ULAs to that of the RIS array as $\xi_{T,\text{RIS}}(t) = \Vert \textbf{d}_\text{RIS} - \textbf{d}_T(t) \Vert$ and $\xi_{R,\text{RIS}}(t) = \Vert \textbf{d}_\text{RIS} - \textbf{d}_R(t) \Vert$, with $\Vert \cdot \Vert$ representing the Frobenius norm. The Fraunhofer distance, which distinguishes between the far-field and near-field ranges in RIS-enabled channel scenarios, can be expressed as \cite{YJPan}
\begin{eqnarray}
L_\text{RIS} = \frac{2 d^2_M \big( (M_x-1)^2 + (M_z-1)^2 \big) }{\lambda}   \, ,
\end{eqnarray}
with $\lambda$ being the wavelength. Due to the dynamic features of the UAV and vehicle, the distance between the UAV/vehicle and the RIS continually vary in the real-time motion stage. This will result in the alternating presence of near-field and far-field ranges \cite{Xiongwcl}, making the conventional methods that rely solely on spherical or planar-wave-based channel models inadequate for assessing the communication system performance for UAV-to-vehicle scenarios incorporating RIS technology. To tackle this challenge, we propose an analytical framework that evenly divides the entire RIS array into $M^\text{sub}_x(t) \times M^\text{sub}_z(t)$ sub-arrays in an evenly distributed manner. The largest sub-array contains $M^\text{sub}_{x,\text{max}}(t) \times M^\text{sub}_{z,\text{max}}(t)$ RIS reflecting elements, as illustrated in Fig. 2. For the convenience of the subsequent derivations, assume that the largest sub-array is square under the premise of acceptable computational complexity, i.e., $M^\text{sub}_{x,\text{max}}(t) = M^\text{sub}_{z,\text{max}}(t)$. In order to ensure that the planar-wave model can be applied to each sub-array, the distances $\xi_{T,\text{RIS}}(t)$ and $\xi_{R,\text{RIS}}(t)$  are required to meet the following constraint condition: \cite{MCui2022}
\begin{align}
\xi&_{T,\text{RIS}}(t) \geq \frac{2 \big( P\delta_T + \sqrt{2}d_M (M^{\text{sub}}_{{x/z},\text{max}}(t)-1)\big)^2} {\lambda}   \, ,
\end{align}
\begin{align}
\xi&_{R,\text{RIS}}(t) \geq \frac{2 \big( Q\delta_R + \sqrt{2}d_M (M^{\text{sub}}_{{x/z},\text{max}}(t)-1 )\big)^2} {\lambda}   \, .
\end{align}
In light of this, we can obtain the constraint of the largest sub-array $M^\text{sub}_{x/z,\text{max}} (t)$ as follows:
\begin{align}
M^\text{sub}_{x/z,\text{max}} (t) \leq \min \Bigg\{ &\underbrace{ \frac{\sqrt{\lambda \xi_{T,\text{RIS}}(t)}}{2 d_M} - \frac{P\delta_T}{\sqrt{2}d_M} + 1 }_{g_1}, \nonumber \\
&\underbrace{ \frac{\sqrt{\lambda \xi_{R,\text{RIS}}(t)}}{2 d_M} - \frac{Q\delta_R}{\sqrt{2}d_M} + 1 }_{g_2} \Bigg\}  \, .
\end{align}
In addition to the above constraint conditions, it is crucial to ensure that the dimension of the RIS panel in each sub-array remains smaller than that of the entire RIS array, that is, $M^\text{sub}_{x/z,\text{max}} (t) \leq M$. Therefore, the $ M^\text{sub}_{x/z,\text{max}} (t)$ also needs to satisfy the following constraint:
\begin{eqnarray}
M^\text{sub}_{x/z,\text{max}} (t) \hspace*{-0.225cm}&=&\hspace*{-0.225cm} \begin{cases} \min \Big \{ \big\lfloor g_1\big\rfloor, \big\lfloor g_2 \big\rfloor, M \Big \}, \\[0.125cm]
\hspace*{2cm} \textrm{if $\min{\{ g_1, g_2 \}} > 1$ }  \\[0.125cm]
1 , \hspace*{1.67cm} \textrm{if $\min{\{ g_1, g_2 \}} \leq 1$ }  \end{cases} \hspace*{-0.4cm} ,
\end{eqnarray}
where $\lfloor x \rfloor$ denotes the operation of the largest integer not greater than $x$. Here, we have
\begin{eqnarray}
M^\text{sub}_{x/z} (t) \hspace*{-0.225cm}&=&\hspace*{-0.225cm} \begin{cases} \frac{M_{x/z} - \text{mod} \big( M_{x/z}, \; M^\text{sub}_{x/z,\text{max}}(t) \big)}{M^\text{sub}_{x/z,\text{max}}(t)} + 1, \\[0.125cm]
\hspace*{1cm} \textrm{if $\text{mod} \big( M_{x/z}, \; M^\text{sub}_{x/z,\text{max}}(t) \big) \neq 0$ }  \\[0.125cm]
M_{x/z} / M^\text{sub}_{x/z,\text{max}} (t) , \\[0.125cm]
\hspace*{1cm} \textrm{if $\text{mod} \big( M_{x/z}, \; M^\text{sub}_{x/z,\text{max}}(t) \big) = 0$ }  \end{cases} \hspace*{-0.4cm} .
\end{eqnarray}
Consequently, for the $(m^\text{sub}_x, m^\text{sub}_z)$-th sub-array, i.e., $m^\text{sub}_x = 1, 2, ..., M^\text{sub}_x(t)$ and $m^\text{sub}_z = 1, 2, ..., M^\text{sub}_z(t)$, their numbers of RIS reflecting elements are represented by
\begin{eqnarray}
M_{m^\text{sub}_{x/z}}(t) = \begin{cases}  M^\text{sub}_{x/z} (t),  \hspace*{0.85cm} \textrm{if $1  \leq  m^\text{sub}_{x/z} < M^{\text{sub}}_{x/z} (t)$ }    \\[0.125cm]    M_{x/z}  -  \big(M^\text{sub}_{x/z} (t) - 1 \big) M^\text{sub}_{x/z} (t) ,  \\[0.125cm]  \hspace*{2.13cm}  \textrm{if  $m^{\text{sub}}_{x/z} = M^{\text{sub}}_{x/z} (t)$ }  \end{cases}  \, \hspace*{-0.4cm} .
\end{eqnarray}
For the convenience of subsequent expressions, define the $m^\text{sub}_{x,z}$ as the $(m^\text{sub}_x, m^\text{sub}_z)$. It should be noted that the sub-array partition framework proposed is primarily designed for near-field communication scenarios. However, when the UAV and vehicle operate in far-field ranges of RIS-enabled wireless channels, planar-wave-based models can be utilized to capture the propagation properties of RIS, eliminating the need to consider sub-array partition. Moreover, it is evident that the parameters of RIS reflecting elements can be set arbitrarily, thus indicating the suitability of the proposed framework for various configurations of RIS arrays and highlighting its universality.


\section{Channel Modeling by Utilizing the Sub-Array Partition Analytical Framework}

In this section, we will model the RIS-enabled UAV-to-vehicle channels by utilizing the above sub-array partition analytical framework. Previous studies, such as \cite{Xiong1} and \cite{Jhwcl}, have derived a complex matrix to represent the physical characteristics of RIS-enabled wireless channels. Each element of this matrix corresponds to the complex channel impulse response (CIR) between a specific pair of transmit-receive antenna arrays in RIS-enabled channels. To effectively capture the sparsity properties of UAV-to-vehicle communication scenarios with RIS, we adopt the complex matrix $\textbf{H}_\text{B} (t, \tau) \in \mathbb{C}^{Q \times P}$ as the beam-domain RIS-enabled UAV-to-vehicle channel model within the proposed framework. It is assumed that the RIS and NLoS propagation links operate independently. Therefore, we have
\begin{eqnarray}\label{H_B_pq_t}
\textbf{H}_\text{B}(t, \tau) = \sqrt{\frac{K}{K+1}} \textbf{H}^\text{RIS}_\text{B} (t, \tau) + \sqrt{\frac{1}{K+1}} \textbf{H}^\text{NLoS}_\text{B} (t, \tau) \, ,
\end{eqnarray}
with $K$ being the Rician factor, defined as the percentage of RIS links to the NLoS links. The $\textbf{H}^\text{RIS}_\text{B} (t, \tau)$ is the RIS propagation components for the proposed beam-domain channel matrix, which can be expressed as
\begin{eqnarray}\label{H_B_RIS_pq_t}
\textbf{H}^\text{RIS}_\text{B} (t, \tau) = \widetilde{\mathbf{V}}^\text{H} \textbf{H}^\text{RIS}_\text{G} (t, \tau) \widetilde{\mathbf{U}}^*   \, ,
\end{eqnarray}
where ($\cdot$)$^*$ stands for the complex conjugate operation. The transform matrix $\widetilde{\mathbf{U}}$ converts the antenna domain channel matrix to the beam domain channel matrix at the UAV side, which is expressed as
\begin{eqnarray}
\widetilde{\mathbf{U}} = \sqrt{\frac{1}{P}} \big[ \, \textbf{a}(\theta_p) \, \big] \in \mathbb{C}^{P \times P} \,  ,
\end{eqnarray}
with $\theta_p = (p-0.5-0.5P)/P$ representing the assigned azimuth spatial frequencies at UAV side, with a range of $[ \, -0.5, 0.5 \, ]$. The transmitted beamforming matrix $\widetilde{\mathbf{V}}$ at the vehicle receiver, as depicted in (\ref{H_B_RIS_pq_t}), is represented by
\begin{eqnarray}
\widetilde{\mathbf{V}} = \sqrt{\frac{1}{Q}} \big[ \, \textbf{a}(\theta_q) \, \big] \in \mathbb{C}^{Q \times Q} \,  ,
\end{eqnarray}
with $\theta_{q} = (q-0.5-0.5Q)/Q$ representing the assigned azimuth spatial frequencies at the vehicle side, with a range of $[ \, -0.5, 0.5 \, ]$. In (11), $\textbf{H}^\text{RIS}_\text{G} (t, \tau)$ denotes the RIS propagation components for the proposed channel matrix based on the geometry-based method, which can be represented by
\begin{align}
&\textbf{H}^\text{RIS}_\text{G} (t, \tau) = h^\text{RIS}_{pq, m^\text{sub}_{x,z}} (t, \tau) \nonumber \\[0.125cm]
& \times \textbf{v}^\text{T} (\theta^\text{azi}_{R,m^\text{sub}_{x,z}}(t), \theta^\text{ver}_{R,m^\text{sub}_{x,z}}(t)) \textbf{u}(\theta^\text{azi}_{T,m^\text{sub}_{x,z}}(t), \theta^\text{ver}_{T,m^\text{sub}_{x,z}}(t)) \, ,
\end{align}
where $\textbf{u}(\theta^\text{azi}_{T,m^\text{sub}_{x,z}}(t), \theta^\text{ver}_{T,m^\text{sub}_{x,z}}(t)) \in \mathbb{C}^{1 \times P}$ denotes the response vector of the UAV antenna array, which can be represented by
\begin{align}
\textbf{u}(\theta^\text{azi}_{T,m^\text{sub}_{x,z}}(t), \theta&^\text{ver}_{T,m^\text{sub}_{x,z}}(t)) = \big[ \, 1 , e^{j2\pi \big( \theta^\text{azi}_{T,m^\text{sub}_{x,z}}(t) + \theta^\text{ver}_{T,m^\text{sub}_{x,z}}(t) \big)} , \, \nonumber \\
&... , e^{j2\pi(P-1)\big(\theta^\text{azi}_{T,m^\text{sub}_{x,z}}(t) + \theta^\text{ver}_{T,m^\text{sub}_{x,z}}(t) \big)} \, \big]  \,  ,
\end{align}
where $j = \sqrt{-1}$ is the imaginary unit. Here, $\theta^\text{azi}_{T,m^\text{sub}_{x,z}}(t)$ and $\theta^\text{ver}_{T,m^\text{sub}_{x,z}}(t)$ account for the spatial frequencies along the azimuth and vertical dimensions, respectively, which can be represented by
\begin{align}
\theta&^\text{azi}_{T,m^\text{sub}_{x,z}}(t) = \nonumber \\[0.125cm]
&\delta_T / \lambda \cos\alpha^\text{ver}_{T,m^\text{sub}_{x,z}}(t) \cos\psi^\text{ver}_T \cos\big( \alpha^\text{azi}_{T,m^\text{sub}_{x,z}}(t) - \psi^\text{azi}_T \big) \, ,
\end{align}
\begin{eqnarray}
\theta^\text{ver}_{T,m^\text{sub}_{x,z}}(t) = \delta_T / \lambda \sin\alpha^\text{ver}_{T,m^\text{sub}_{x,z}}(t) \sin\psi^\text{ver}_T \, .
\end{eqnarray}
Furthermore, $\textbf{v} (\theta^\text{azi}_{R,m^\text{sub}_{x,z}}(t), \theta^\text{ver}_{R,m^\text{sub}_{x,z}}(t)) \in \mathbb{C}^{1 \times Q}$ is the response vector of the vehicle antenna array, which can be expressed as
\begin{align}
\textbf{v} ( \theta^\text{azi}_{R,m^\text{sub}_{x,z}}(t),\theta&^\text{ver}_{R,m^\text{sub}_{x,z}}(t) ) = \big[ \, 1 , e^{j2\pi \big( \theta^\text{azi}_{R,m^\text{sub}_{x,z}}(t) + \theta^\text{ver}_{R,m^\text{sub}_{x,z}}(t) \big)} , \, \nonumber \\
&... , e^{j2\pi(Q-1)\big(\theta^\text{azi}_{R,m^\text{sub}_{x,z}}(t) + \theta^\text{ver}_{R,m^\text{sub}_{x,z}}(t) \big)} \, \big]  \,  .
\end{align}
Here, $\theta^\text{azi}_{R,m^\text{sub}_{x,z}}(t)$ and $\theta^\text{ver}_{R,m^\text{sub}_{x,z}}(t)$ are respectively the spatial frequencies at the vehicle side along the azimuth and vertical dimensions, which can be expressed as
\begin{align}
\theta&^\text{azi}_{R,m^\text{sub}_{x,z}}(t) = \nonumber \\[0.125cm]
&\delta_R / \lambda \cos\alpha^\text{ver}_{R,m^\text{sub}_{x,z}}(t) \cos\psi^\text{ver}_R \cos\big( \alpha^\text{azi}_{R,m^\text{sub}_{x,z}}(t) - \psi^\text{azi}_R \big) \, ,
\end{align}
\begin{eqnarray}
\theta^\text{ver}_{R,m^\text{sub}_{x,z}}(t) = \delta_R / \lambda \sin\alpha^\text{ver}_{R,m^\text{sub}_{x,z}}(t) \sin\psi^\text{ver}_R \, .
\end{eqnarray}
In (14), $h^\text{RIS}_{pq, m^\text{sub}_{x,z}} (t, \tau)$ is the complex CIR of RIS propagation component for the $(p, q)$-th antenna pair through the $m^\text{sub}_{x,z}$-th RIS sub-array, which is denoted by
\allowdisplaybreaks[4]
\begin{align}
h^\text{RIS}_{pq, m^\text{sub}_{x,z}} &(t, \tau) = \nonumber \\[0.125cm]
&h^\text{RIS}_{pq, m^\text{sub}_{x,z}} (t) \delta \big( \tau - \frac{\xi_{T,m^\text{sub}_{x,z}}(t) + \xi_{R,m^\text{sub}_{x,z}}(t)}{c} \big) \, ,
\end{align}
where $c$ denotes the light speed. The $\xi_{T,m^\text{sub}_{x,z}}(t)$ and $\xi_{R,m^\text{sub}_{x,z}}(t)$ are the real-time propagation distances from the midpoints of the UAV and vehicle ULAs to the reflecting units in the $m^\text{sub}_{x,z}$-th RIS sub-array, respectively. These distances can be derived using the formulas $\xi_{T,m^\text{sub}_{x,z}}(t) = \Vert \textbf{d}_{m^\text{sub}_{x,z}} - \textbf{d}_T(t) \Vert$ and $\xi_{R,m^\text{sub}_{x,z}}(t) = \Vert \textbf{d}_{m^\text{sub}_{x,z}} - \textbf{d}_R(t) \Vert$. Here, $\textbf{d}_{m^\text{sub}_{x,z}}$ represents the distance vector from the origin of the global coordinate system to the center of the $m^\text{sub}_{x,z}$-th sub-array in the RIS, given as $\textbf{d}_{m^\text{sub}_{x,z}} = [ \, x_{m^\text{sub}_{x,z}}, y_{m^\text{sub}_{x,z}}, z_{m^\text{sub}_{x,z}} \, ]^\text{T}$. Additionally, the channel coefficient $h^\text{RIS}_{pq, m^\text{sub}_{x,z}} (t)$ of the RIS propagation component for the $(p, q)$-th antenna pair through the $m^\text{sub}_{x,z}$-th RIS sub-array is expressed as
\allowdisplaybreaks[4]
\begin{eqnarray}
\!\!\!\!h^\text{RIS}_{pq, m^\text{sub}_{x,z}} (\hspace*{-0.425cm}&t&\hspace*{-0.425cm}) = \sum_{m^\text{sub}_x = 1}^{ M^\text{sub}_x(t)} \sum_{m^\text{sub}_z = 1}^{ M^\text{sub}_z(t)} \chi_{m^\text{sub}_{x,z}}(t) e^{j \varphi_{m^\text{sub}_{x,z}}(t)}  \nonumber \\ [0.125cm]
&\times&\hspace*{-0.225cm}  e^{- j\frac{2\pi}{\lambda} \big( \xi_{T,m^\text{sub}_{x,z}}(t) + \xi_{R,m^\text{sub}_{x,z}}(t) \big) }  \nonumber \\ [0.125cm]
&\times&\hspace*{-0.225cm}  e^{ j\frac{2\pi}{\lambda} \big< \textbf{e}_{T,m^\text{sub}_{x,z}}(t), \, \textbf{d}_{T,p} \big> + j\frac{2\pi}{\lambda} \big< \textbf{e}_{R,m^\text{sub}_{x,z}}(t), \, \textbf{d}_{R,q} \big> } \nonumber \\ [0.125cm]
&\times&\hspace*{-0.225cm}  e^{ j\frac{2\pi}{\lambda} \big< \textbf{v}_Tt, \, \textbf{e}_{T,m^\text{sub}_{x,z}}(t) \big> } e^{ j\frac{2\pi}{\lambda} \big< \textbf{v}_Rt, \, \textbf{e}_{R,m^\text{sub}_{x,z}}(t) \big> }  \, ,
\end{eqnarray}
Here, $\big<\cdot, \cdot\big>$ is the vector dot product, $\chi_{m^\text{sub}_{x,z}}(t)$ and $\varphi_{m^\text{sub}_{x,z}}(t)$ are respectively the real-time regulation amplitude and phase of the reflecting units in the $m^\text{sub}_{x,z}$-th RIS sub-array. Moreover, $\textbf{v}_T$ and $\textbf{v}_R$ are the velocity vectors of the UAV and vehicle, respectively. These velocity vectors can be expressed as $\textbf{v}_T = v_T [\, \cos\eta^\text{ver}_T\cos\eta^\text{azi}_T, \cos\eta^\text{ver}_T\sin\eta^\text{azi}_T, \sin\eta^\text{ver}_T \,]^\text{T}$ and $\textbf{v}_R = v_R [\, \cos\eta^\text{azi}_R, \sin\eta^\text{azi}_R, 0 \,]^\text{T}$. In addition, $\textbf{e}_{T, {m^\text{sub}_{x,z}}}(t)$ and $\textbf{e}_{R, {m^\text{sub}_{x,z}}}(t)$ are respectively the real-time steering vectors of the RIS propagation component from the UAV and vehicle to the reflecting units in the $m^\text{sub}_{x,z}$-th RIS sub-array, which are expressed as
\allowdisplaybreaks[4]
\begin{eqnarray}
\textbf{e}_{T/R, m^\text{sub}_{x,z}}(t) = \left[ \begin{array}{ccc} \, \cos\alpha^\text{ver}_{T/R,m^\text{sub}_{x,z}}(t)\cos\alpha^\text{azi}_{T/R,m^\text{sub}_{x,z}}(t) \, \\ \cos\alpha^\text{ver}_{T/R,m^\text{sub}_{x,z}}(t) \sin\alpha^\text{azi}_{T/R,m^\text{sub}_{x,z}}(t) \\ \sin\alpha^\text{ver}_{T/R,m^\text{sub}_{x,z}}(t) \end{array} \right]   \, ,
\end{eqnarray}
where $\alpha^\text{azi}_{T,m^\text{sub}_{x,z}}(t)$ and $\alpha^\text{ver}_{T,m^\text{sub}_{x,z}}(t)$ respectively stand for the real-time angular parameters along the azimuth and vertical dimensions from the UAV to the reflecting units in the $m^\text{sub}_{x,z}$-th RIS sub-array, which can be expressed as
\begin{eqnarray}
\alpha^\text{azi}_{T,m^\text{sub}_{x,z}}(t) = \arctan \frac{ y_{m^\text{sub}_{x,z}} - d_{T,y}(t) }{ x_{m^\text{sub}_{x,z}} - d_{T,x}(t) } ,
\end{eqnarray}
\begin{align}
\alpha&^\text{ver}_{T,m^\text{sub}_{x,z}}(t) = \nonumber \\
&\arctan \frac{ z_{m^\text{sub}_{x,z}} - d_{T,z}(t) }{ \sqrt{ \big( x_{m^\text{sub}_{x,z}} - d_{T,x}(t) \big)^2 + \big( y_{m^\text{sub}_{x,z}} - d_{T,y}(t) \big)^2 } }  .
\end{align}
Furthermore, $\alpha^\text{azi}_{R,m^\text{sub}_{x,z}}(t)$ and $\alpha^\text{ver}_{R,m^\text{sub}_{x,z}}(t)$ are respectively the real-time angular parameters along the azimuth and vertical dimensions from the reflecting units in the $m^\text{sub}_{x,z}$-th RIS sub-array to the vehicle, which can be expressed as
\begin{eqnarray}
\alpha^\text{azi}_{R,m^\text{sub}_{x,z}}(t) = \arctan \frac{ y_{m^\text{sub}_{x,z}} - d_{R,y}(t) }{ x_{m^\text{sub}_{x,z}} - d_{R,x}(t) } ,
\end{eqnarray}
\begin{align}
\alpha&^\text{ver}_{R,m^\text{sub}_{x,z}}(t) = \nonumber \\[0.125cm]
&\arctan \frac{ z_{m^\text{sub}_{x,z}} }{ \sqrt{ \big( x_{m^\text{sub}_{x,z}} - d_{R,x}(t) \big)^2 + \big( y_{m^\text{sub}_{x,z}} - d_{R,y}(t) \big)^2 } } \, .
\end{align}
Therefore, the complex CIR of RIS propagation component for the $(p, q)$-th antenna pair through the $m^\text{sub}_{x,z}$-th RIS sub-array under the beam-domain method can be expressed as
\allowdisplaybreaks[4]
\begin{eqnarray}
h^\text{RIS}_{\text{B}, (pq, m^\text{sub}_{x,z})} (\hspace*{-0.425cm}&t&\hspace*{-0.425cm},\tau) = \sqrt{\frac{1}{PQ}} \, \sum_{m^\text{sub}_x = 1}^{ M^\text{sub}_{x}(t)} \sum_{m^\text{sub}_z = 1}^{ M^\text{sub}_{z}(t)} \chi_{m^\text{sub}_{x,z}}(t) e^{j \varphi_{m^\text{sub}_{x,z}}(t)} \nonumber \\
&\times&\hspace*{-0.225cm}  e^{- j\frac{2\pi}{\lambda} \big( \xi_{T,m^\text{sub}_{x,z}}(t) + \xi_{R,m^\text{sub}_{x,z}}(t) \big) }  \nonumber \\ [0.125cm]
&\times&\hspace*{-0.225cm}  e^{ j\frac{2\pi}{\lambda} \big< \textbf{v}_Tt, \, \textbf{e}_{T,m^\text{sub}_{x,z}}(t) \big> } e^{ j\frac{2\pi}{\lambda} \big< \textbf{v}_Rt, \, \textbf{e}_{R,m^\text{sub}_{x,z}}(t) \big> }  \nonumber \\ [0.125cm]
&\times&\hspace*{-0.225cm} \sum_{\varepsilon = 0}^{P-1} e^{j 2\pi \varepsilon \big( \theta^\text{ver}_{T, m^\text{sub}_{x,z}}(t) + \theta^\text{azi}_{T, m^\text{sub}_{x,z}}(t) - \theta_p\big) } \nonumber \\ [0.125cm]
&\times&\hspace*{-0.225cm} \sum_{\kappa = 0}^{Q-1} e^{j 2\pi \kappa \big( \theta^\text{ver}_{R, m^\text{sub}_{x,z}}(t) + \theta^\text{azi}_{R, m^\text{sub}_{x,z}}(t) - \theta_q\big) }  \nonumber \\ [0.125cm]
&\times&\hspace*{-0.225cm} \delta \big( \tau - \frac{\xi_{T,m^\text{sub}_{x,z}}(t) + \xi_{R,m^\text{sub}_{x,z}}(t)}{c} \big) ,
\end{eqnarray}
with $\theta_p = (p-0.5-0.5P)/P$ and $\theta_q = (q-0.5-0.5Q)/Q$ representing the assigned azimuth spatial frequencies at UAV and vehicle sides, respectively, with a range of $[ \, -0.5, 0.5 \, ]$.

In (\ref{H_B_pq_t}), $\textbf{H}^\text{NLoS}_\text{B} (t, \tau)$ is the NLoS propagation components for the proposed beam-domain channel matrix, which is represented by
\begin{align}\label{H_B_NLoS_pq_t}
\textbf{H}&^\text{NLoS}_\text{B} (t, \tau) = \widetilde{\mathbf{V}}^\text{H} \textbf{H}^\text{NLoS}_\text{G} (t, \tau) \widetilde{\mathbf{U}}^* \nonumber \\[0.125cm]
&= h^\text{NLoS}_{pq} (t, \tau) \textbf{v}^\text{T} (\theta^\text{azi}_{R,n_{\ell}}(t), \theta^\text{ver}_{R,n_{\ell}}(t)) \textbf{u}(\theta^\text{azi}_{T,n_{\ell}}(t), \theta^\text{ver}_{T,n_{\ell}}(t))
\, ,
\end{align}
where $\textbf{u} (\theta^\text{azi}_{R,n_{\ell}}(t), \theta^\text{ver}_{R,n_{\ell}}(t))$ and $\textbf{v}(\theta^\text{azi}_{T,n_{\ell}}(t), \theta^\text{ver}_{T,n_{\ell}}(t))$ are the response vectors of the NLoS propagation link at the UAV and vehicle sides, respectively, which is represented by
\begin{align}
\textbf{u}(\theta^\text{azi}_{T,n_{\ell}}(t), \theta^\text{ver}_{T,n_{\ell}}&(t)) = \big[ \, 1 , e^{j2\pi \big( \theta^\text{azi}_{T,n_{\ell}}(t) + \theta^\text{ver}_{T,n_{\ell}}(t) \big)} ,... \, \nonumber \\[0.125cm]
&, e^{j2\pi(P-1)\big(\theta^\text{azi}_{T,n_{\ell}}(t) + \theta^\text{ver}_{T,n_{\ell}}(t) \big)} \, \big]  \,  ,
\end{align}
\begin{align}
\textbf{v} ( \theta^\text{azi}_{R,n_{\ell}}(t),\theta^\text{ver}_{R,n_{\ell}}&(t) ) = \big[ \, 1 , e^{j2\pi \big( \theta^\text{azi}_{R,n_{\ell}}(t) + \theta^\text{ver}_{R,n_{\ell}}(t) \big)} ,... \, \nonumber \\[0.125cm]
&, e^{j2\pi(Q-1)\big(\theta^\text{azi}_{R,n_{\ell}}(t) + \theta^\text{ver}_{R,n_{\ell}}(t) \big)} \, \big]  \,  ,
\end{align}
in which $\theta^\text{azi}_{T,n_{\ell}}(t)$ and $\theta^\text{azi}_{R,n_{\ell}}(t)$ respectively stand for the spatial frequency along the azimuth dimension associated with the path from the midpoints of UAV's ULA and vehicle's ULA to the $n_{\ell}$-th path within the $n$-th cluster, which are expressed as
\begin{align}
&\theta^\text{azi}_{T/R,n_{\ell}}(t) = \nonumber \\[0.125cm]
&\delta_{T/R} / \lambda \cos\alpha^\text{ver}_{T/R,n_{\ell}}(t) \cos\psi^\text{ver}_{T/R} \cos\big( \alpha^\text{azi}_{T/R,n_{\ell}}(t) - \psi^\text{azi}_{T/R} \big) \, ,
\end{align}
where $\theta^\text{ver}_{T,n_{\ell}}(t)$ and $\theta^\text{ver}_{R,n_{\ell}}(t)$ are respectively the spatial frequency along the vertical dimension associated with the path from the midpoints of UAV¡¯s ULA and vehicle¡¯s ULA to the $n_{\ell}$-th path within the $n$-th cluster, which are expressed as
\begin{eqnarray}
\theta^\text{ver}_{T/R,n_{\ell}}(t) = \delta_{T/R} / \lambda \sin\alpha^\text{ver}_{T/R,n_{\ell}}(t) \sin\psi^\text{ver}_{T/R} \, .
\end{eqnarray}
In (29), $h^\text{NLoS}_{pq}(t,\tau)$ is the complex CIR of NLoS propagation component for the $(p, q)$-th antenna pair through the NLoS propagations, which is denoted by
\allowdisplaybreaks[4]
\begin{eqnarray}
h^\text{NLoS}_{pq} (t,\tau) = h^\text{NLoS}_{pq} (t)  \delta \big( \tau - \frac{\xi_{T,\text{cluster}}(t) + \xi_{R,\text{cluster}}(t)}{c} \big) \, ,
\end{eqnarray}
where $\xi_{T,\text{cluster}}(t)$ and $\xi_{R,\text{cluster}}(t)$ denote the real-time propagation distances from the midpoints of the UAV ULA and vehicle ULA to that of cluster, respectively, they are $\xi_{T,\text{cluster}}(t) = \Vert \textbf{d}_\text{cluster} - \textbf{d}_T(t) \Vert$ and $\xi_{R,\text{cluster}}(t) = \Vert \textbf{d}_\text{cluster} - \textbf{d}_R(t) \Vert$, with $\textbf{d}_\text{cluster}$ representing the distance vector from the origin of the global coordinate system to the midpoint of the cluster, i.e., $\textbf{d}_\text{cluster} = [ \, x_\text{cluster}, y_\text{cluster}, z_\text{cluster} \, ]^\text{T}$. Furthermore, the channel coefficient $h^\text{NLoS}_{pq} (t)$ can be written by
\begin{eqnarray}
h^\text{NLoS}_{pq} (t) \hspace*{-0.225cm}&=&\hspace*{-0.225cm} \sum_{n_\ell = 1}^{n_L} e^{j\varphi_0 - j\frac{2\pi}{\lambda} \big( \xi_{T,n_\ell}(t) + \xi_{R,n_\ell}(t) \big) }  \nonumber \\ [0.125cm]
&\times&\hspace*{-0.225cm}  e^{ j\frac{2\pi}{\lambda} \big< \textbf{e}_{T,n_\ell}(t), \, \textbf{d}_{T,p} \big> + j\frac{2\pi}{\lambda} \big< \textbf{e}_{R,n_\ell}(t), \, \textbf{d}_{R,q} \big> } \nonumber \\ [0.125cm]
&\times&\hspace*{-0.225cm}  e^{ j\frac{2\pi}{\lambda} \big< \textbf{v}_Tt, \, \textbf{e}_{T,n_\ell}(t) \big> } e^{ j\frac{2\pi}{\lambda} \big< \textbf{v}_Rt, \, \textbf{e}_{R,n_\ell}(t) \big> }   \, ,
\end{eqnarray}
with $n_L$ representing the number of paths within the $n$-th cluster, $\xi_{T,n_\ell}(t)$ and $\xi_{R,n_\ell}(t)$ are the real-time propagation distances from the midpoints of the UAV ULA and vehicle ULA to the $n_\ell$-th ray within the cluster, respectively, they are $\xi_{T,n_\ell}(t) = \Vert \textbf{d}_{n_\ell} - \textbf{d}_T(t) \Vert$ and $\xi_{R,n_\ell}(t) = \Vert \textbf{d}_{n_\ell} - \textbf{d}_R(t) \Vert$, with $\textbf{d}_{n_\ell}$ representing the distance vector from the origin of the global coordinate system to the $n_\ell$-th scatterer within the cluster, i.e., $\textbf{d}_{n_\ell} = [ \, x_{n_\ell}, y_{n_\ell}, z_{n_\ell} \, ]^\text{T}$. Besides, $\textbf{e}_{T,n_\ell}(t)$ and $\textbf{e}_{R,n_\ell}(t)$ are respectively the unit directional vectors from UAV and vehicle to the $n_\ell$-th ray within the cluster, which are expressed as
\allowdisplaybreaks[4]
\begin{eqnarray}
\textbf{e}_{T/R,n_\ell}(t) = \left[ \begin{array}{ccc} \, \cos\alpha^\text{ver}_{T/R,n_\ell}(t)\cos\alpha^\text{azi}_{T/R,n_\ell}(t) \, \\ \cos\alpha^\text{ver}_{T/R,n_\ell}(t)\sin\alpha^\text{azi}_{T/R,n_\ell}(t) \\ \sin\alpha^\text{ver}_{T/R,n_\ell}(t) \end{array} \right]   \, ,
\end{eqnarray}
where $\alpha^\text{azi}_{T,n_\ell}(t)$ and $\alpha^\text{ver}_{T,n_\ell}(t)$ respectively stand for the real-time angular parameters along the azimuth and vertical dimensions at the UAV side, which can be expressed as
\begin{eqnarray}
\alpha^\text{azi}_{T,n_\ell}(t) = \arctan \frac{ y_{n_\ell} - d_{T,y}(t) }{ x_{n_\ell} - d_{T,x}(t) } ,
\end{eqnarray}
\begin{align}
\alpha&^\text{ver}_{T,n_\ell}(t) = \nonumber \\
&\arctan \frac{ z_{n_\ell} - d_{T,z}(t) }{ \sqrt{ \big( x_{n_\ell} - d_{T,x}(t) \big)^2 + \big( y_{n_\ell} - d_{T,y}(t) \big)^2 } }  .
\end{align}
Furthermore, $\alpha^\text{azi}_{R,n_\ell}(t)$ and $\alpha^\text{ver}_{R,n_\ell}(t)$ are respectively the real-time angular parameters along the azimuth and vertical dimensions at the vehicle side, which can be expressed as
\begin{eqnarray}
\alpha^\text{azi}_{R,n_\ell}(t) = \arctan \frac{ y_{n_\ell} - d_{R,y}(t) }{ x_{n_\ell} - d_{R,x}(t) } ,
\end{eqnarray}
\begin{align}
\alpha&^\text{ver}_{R,n_\ell}(t) = \nonumber \\[0.125cm]
&\arctan \frac{ z_{n_\ell} }{ \sqrt{ \big( x_{n_\ell} - d_{R,x}(t) \big)^2 + \big( y_{n_\ell} - d_{R,y}(t) \big)^2 } }  \, .
\end{align}
Therefore, the channel coefficient of NLoS propagation component for the $(p, q)$-th antenna pair based on the beam-domain method can be expressed as
\allowdisplaybreaks[4]
\begin{eqnarray}
h^\text{NLoS}_{\text{B},pq} (t,\tau) \hspace*{-0.225cm}&=&\hspace*{-0.225cm} \sqrt{\frac{1}{PQ}} \, \sum_{n_\ell = 1}^{n_L} e^{- j\frac{2\pi}{\lambda} \big( \xi_{T,n_\ell}(t) + \xi_{R,n_\ell}(t) \big) }  \nonumber \\ [0.125cm]
&\times&\hspace*{-0.225cm}  e^{ j\frac{2\pi}{\lambda} \big< \textbf{e}_{T,n_\ell}(t), \, \textbf{d}_{T,p} \big> + j\frac{2\pi}{\lambda} \big< \textbf{e}_{R,n_\ell}(t), \, \textbf{d}_{R,q} \big> } \nonumber \\ [0.125cm]
&\times&\hspace*{-0.225cm}  e^{ j\frac{2\pi}{\lambda} \big< \textbf{v}_Tt, \, \textbf{e}_{T,n_\ell}(t) \big> } e^{ j\frac{2\pi}{\lambda} \big< \textbf{v}_Rt, \, \textbf{e}_{R,n_\ell}(t) \big> }   \nonumber \\ [0.125cm]
&\times&\hspace*{-0.225cm} \sum_{\varepsilon = 0}^{P-1} e^{j 2\pi \varepsilon \big( \theta^\text{ver}_{T,n_\ell}(t) + \theta^\text{azi}_{T,n_\ell}(t) - \theta_p\big) } \nonumber \\ [0.125cm]
&\times&\hspace*{-0.225cm} \sum_{\kappa = 0}^{Q-1} e^{j 2\pi \kappa \big( \theta^\text{ver}_{R,n_\ell}(t) + \theta^\text{azi}_{R,n_\ell}(t) - \theta_q\big) } \nonumber \\ [0.125cm]
&\times&\hspace*{-0.225cm} \delta \big( \tau - \frac{\xi_{T,\text{cluster}}(t) + \xi_{R,\text{cluster}}(t)}{c} \big) \,  .
\end{eqnarray}
It shows that the complex CIRs of the RIS and NLoS components are partitioned into different beams, which can be used to efficiently capture the sparse features of RIS-enabled UAV-to-vehicle channels. Obviously, the proposed channel modeling method differs significantly from the geometry-based channel modeling method. Hence, a detailed comparison between the two methods will be presented in the following sections.

\section{Propagation Statistics of the Developed Channel Model}

This section aims to analyze the propagation statistics of UAV-to-vehicle channel models, specifically focusing on spatial CCFs, temporal ACFs, frequency CFs, and channel capacities.

\subsection{Spatial-Temporal Correlation Functions}

The RIS-enabled UAV-to-vehicle channel model establishes that the correlation properties of any two arbitrary CIRs, namely $h_{\text{B},pq} (t,\tau)$ and $h_{\text{B},p'q'} (t,\tau)$, are solely determined by the correlation properties of $h_{\text{B},pq} (t)$ and $h_{\text{B},p'q'} (t)$. Here, $p' = 1,2,...,P$ and $q' = 1,2,...,Q$. Under the assumption that there are no interdependencies between various propagation delays and types of propagation paths, we then have
\allowdisplaybreaks[4]
\begin{align}
\rho_{\text{B},(pq,p'q')} &(t, \Delta t) = \nonumber \\[0.125cm]
&\frac{\mathbb{E}\big[h_{\text{B},pq}(t) h^\ast_{\text{B},p'q'}(t + \Delta t) \big]}{\sqrt{\mathbb{E}\big[\vert h_{\text{B},pq}(t) \vert^2 \big] \mathbb{E}\big[\vert h_{\text{B},p'q'}(t + \Delta t) \vert^2 \big] }}               \,\,   ,
\end{align}
where $\Delta t$ denotes the time difference. By substituting Eq. (28) into Eq. (42), we can express the spatial-temporal correlation function between the $(p, q)$-th and the $(p', q')$-th antenna pairs of the RIS components as follows:
\allowdisplaybreaks[4]
\begin{align}
\rho&^\text{RIS}_{\text{B},(pq,p'q')} t, \Delta t) = \frac{K}{K + 1} \mathbb{E}\Big\{  \sum_{m^\text{sub}_x = 1}^{ M^\text{sub}_x(t)} \sum_{m^\text{sub}_z = 1}^{ M^\text{sub}_z(t) } \nonumber \\
&\times  \chi_{m^\text{sub}_{x,z}} (t) \chi_{m^\text{sub}_{x,z}} (t + \Delta t) e^{j \big( \varphi_{m^\text{sub}_{x,z}}(t) - \varphi_{m^\text{sub}_{x,z}}(t + \Delta t)   \big)  } \nonumber \\
&\times  e^{j \frac{2\pi}{\lambda}\big(\xi_{T, m^\text{sub}_{x,z}}(t) - \xi_{T, m^\text{sub}_{x,z}}(t + \Delta t) + \xi_{R, m^\text{sub}_{x,z}}(t) - \xi_{R,m^\text{sub}_{x,z}}(t + \Delta t)  \big) }     \nonumber \\
&\times  e^{ j\frac{2\pi}{\lambda} \big< \textbf{e}_{T,m^\text{sub}_{x,z}}(t), \, \textbf{d}_{T,p} \big> } e^{ j\frac{2\pi}{\lambda} \big< \textbf{e}_{R,,m^\text{sub}_{x,z}}(t), \, \textbf{d}_{R,q} \big> }     \nonumber \\
&\times  e^{ -j\frac{2\pi}{\lambda} \big< \textbf{e}_{T,m^\text{sub}_{x,z}}(t+ \Delta t), \, \textbf{d}_{T,p} \big> } e^{ -j\frac{2\pi}{\lambda} \big< \textbf{e}_{R,m^\text{sub}_{x,z}}(t+ \Delta t), \, \textbf{d}_{R,q} \big> }     \nonumber \\
&\times  e^{ j\frac{2\pi}{\lambda} \big< \textbf{v}_Tt, \, \textbf{e}_{T,m^\text{sub}_{x,z}}(t) \big> } e^{ j\frac{2\pi}{\lambda} \big< \textbf{v}_Rt, \, \textbf{e}_{R,m^\text{sub}_{x,z}}(t) \big> }  \nonumber \\
&\times  e^{ -j\frac{2\pi}{\lambda} \big< \textbf{v}_T \cdot (t+ \Delta t), \, \textbf{e}_{T,m^\text{sub}_{x,z}}(t+ \Delta t) \big> } \nonumber \\
&\times  e^{ -j\frac{2\pi}{\lambda} \big< \textbf{v}_R \cdot (t+ \Delta t), \, \textbf{e}_{R,m^\text{sub}_{x,z}}(t+ \Delta t) \big> } \nonumber \\
&\times \sum_{\varepsilon = 0}^{P-1} e^{j 2\pi \varepsilon \big( \theta^\text{ver}_{T, m^\text{sub}_{x,z}}(t) + \theta^\text{azi}_{T, m^\text{sub}_{x,z}}(t) - \theta_p\big)}  \nonumber \\
&\times \sum_{\kappa = 0}^{Q-1} e^{j 2\pi \kappa \big( \theta^\text{ver}_{R, m^\text{sub}_{x,z}}(t) + \theta^\text{azi}_{R, m^\text{sub}_{x,z}}(t) - \theta_q\big)}  \nonumber \\
&\times \sum_{\varepsilon = 0}^{P-1} e^{-j 2\pi \varepsilon \big( \theta^\text{ver}_{T, m^\text{sub}_{x,z}}(t+\Delta t) + \theta^\text{azi}_{T, m^\text{sub}_{x,z}}(t+\Delta t) - \theta_p \big)}  \nonumber \\
&\times \sum_{\kappa = 1}^{Q} e^{-j 2\pi \kappa \big( \theta^\text{ver}_{R, m^\text{sub}_{x,z}}(t+\Delta t) + \theta^\text{azi}_{R, m^\text{sub}_{x,z}}(t+\Delta t) - \theta_q\big) }  \Big\}  \, .
\end{align}
By substituting Eq. (41) into Eq. (42), the spatial-temporal correlation function between the $(p, q)$-th and the $(p', q')$-th antenna pairs of the NLoS components can be written by
\allowdisplaybreaks[4]
\begin{align}
\rho&^\text{NLoS}_{\text{B},(pq,p'q')} t, \Delta t) = \frac{1}{K + 1} \mathbb{E}\Big\{  \sum_{n_\ell = 1}^{n_L}    \nonumber \\
&\times e^{- j\frac{2\pi}{\lambda} \big( \xi_{T,n_\ell}(t) + \xi_{R,n_\ell}(t) \big) - \xi_{T,n_\ell}(t+\Delta t) - \xi_{R,n_\ell}(t+\Delta t) \big) }   \nonumber \\
&\times e^{ j\frac{2\pi}{\lambda} \big< \textbf{e}_{T,n_\ell}(t), \, \textbf{d}_{T,p} \big> } e^{ j\frac{2\pi}{\lambda} \big< \textbf{e}_{R,n_\ell}(t), \, \textbf{d}_{R,q} \big> }     \nonumber \\
&\times e^{ -j\frac{2\pi}{\lambda} \big< \textbf{e}_{T,n_\ell}(t+ \Delta t), \, \textbf{d}_{T,p} \big> } e^{ -j\frac{2\pi}{\lambda} \big< \textbf{e}_{R,n_\ell}(t+ \Delta t), \, \textbf{d}_{R,q} \big> }     \nonumber \\
&\times e^{ j\frac{2\pi}{\lambda} \big< \textbf{v}_Tt, \, \textbf{e}_{T,n_\ell}(t) \big> } e^{ j\frac{2\pi}{\lambda} \big< \textbf{v}_Rt, \, \textbf{e}_{R,n_\ell}(t) \big> }  \nonumber \\
&\times e^{ -j\frac{2\pi}{\lambda} \big< \textbf{v}_T \cdot (t+ \Delta t), \, \textbf{e}_{T,n_\ell}(t+ \Delta t) \big> }  \nonumber \\
&\times e^{ -j\frac{2\pi}{\lambda} \big< \textbf{v}_R \cdot (t+ \Delta t), \, \textbf{e}_{R,n_\ell}(t+ \Delta t) \big> }  \nonumber \\
&\times \sum_{\varepsilon = 0}^{P-1} e^{j 2\pi \varepsilon \big( \theta^\text{ver}_{T,n_\ell}(t) + \theta^\text{azi}_{T,n_\ell}(t) - \theta_p\big)}  \nonumber \\
&\times \sum_{\kappa = 0}^{Q-1} e^{j 2\pi \kappa \big( \theta^\text{ver}_{R,n_\ell}(t) + \theta^\text{azi}_{R,n_\ell}(t) - \theta_q\big)}  \nonumber \\
&\times \sum_{\varepsilon = 1}^{P} e^{-j 2\pi \varepsilon \big( \theta^\text{ver}_{T,n_\ell}(t+\Delta t) + \theta^\text{azi}_{T,n_\ell}(t+\Delta t) - \theta_p \big)}  \nonumber \\
&\times \sum_{\kappa = 0}^{Q-1} e^{-j 2\pi \kappa \big( \theta^\text{ver}_{R,n_\ell}(t+\Delta t) + \theta^\text{azi}_{R,n_\ell}(t+\Delta t) - \theta_q\big) }  \Big\}  \, .
\end{align}
The dependence of the spatial-temporal correlation functions on the values of $p$, $q$, $p'$, and $q'$ suggests that the wide-sense stationary assumption in the array domain is invalid. Moreover, the real-time angular parameters and path lengths are correlated with the motion characteristics of the UAV and vehicle, indicating the time non-stationarity of the proposed channel model. Additionally, when we set $p = p'$ and $q = q'$, Eqs. (43)-(44) can be derived as the channel temporal ACFs for RIS-enabled UAV-to-vehicle communication systems. The obtained correlation functions possess the ability to measure the temporal correlation characteristics of the propagation links \cite{3GPP}.

\subsection{Frequency CFs}

By utilizing the Fourier transforms of $h^\text{RIS}_{\text{B}, (pq, m^\text{sub}_{x,z})} (t,\tau)$ and $h^\text{NLoS}_{\text{B},pq} (t,\tau)$ to establish the relationships with the propagation delay $\tau$, we can express the time-invariant transfer function of the developed channel model as follows:
\begin{align}
H&_{pq}(t, f) = \nonumber \\[0.125cm]
&\hspace*{0.45cm}\sqrt{\frac{K}{K+1}} h^\text{RIS}_{\text{B}, (pq, m^\text{sub}_{x,z})}(t) e^{-j 2\pi f \big( \xi_{T,m^\text{sub}_{x,z}}(t) + \xi_{R,m^\text{sub}_{x,z}}(t) \big)/c } \nonumber \\[0.125cm]
&+ \sqrt{\frac{1}{K+1}} h^\text{cluster}_{\text{B},pq}(t) e^{-j 2\pi f \big( \xi_{T,\text{cluster}}(t) + \xi_{R,\text{cluster}}(t) \big)/c } \,   ,
\end{align}
Then, we can represent the frequency CF for the propagation link $(p, q)$-th antenna pair as follows:
\begin{align}
\rho_{H_{\text{B},pq}} &(t, \Delta f) = \nonumber \\[0.125cm]
&\frac{\mathbb{E}\big[ \, H_{\text{B},pq}(t) H^\ast_{\text{B},pq}(f + \Delta f) \, \big]}{\sqrt{\mathbb{E}\big[ \, \vert H_{\text{B},pq}(t) \vert^2 \, \big] \mathbb{E}\big[ \, \vert H_{\text{B},pq}(f + \Delta f) \vert^2 \, \big] }}           \,    ,
\end{align}
with $\Delta f$ being the frequency separation. By substituting the expression in Eq. (45) into Eq. (46), we can write the frequency CF of the RIS-enabled UAV-to-vehicle channel model as follows:
\begin{align}
\rho_{H_{\text{B},pq}} &(t, \Delta f) = \frac{K}{K+1} e^{-j 2\pi \Delta f \big( \xi_{T,\text{RIS}}(t) + \xi_{R,\text{RIS}}(t) \big)/c } \nonumber \\[0.125cm]
&+ \frac{1}{K+1} \sum_{\ell=1}^{L} e^{-j 2\pi \Delta f \big( \xi_{T,\text{cluster}}(t) + \xi_{R,\text{cluster}}(t) \big)/c }     \, .
\end{align}
It shows that the channel frequency CFs for RIS-enabled UAV-to-vehicle communications are dependent on the frequency separation $\Delta f$ and the motion time $t$, but they are not influenced by the specific frequency $f$. This observation clearly indicates that the RIS-enabled UAV-to-vehicle channel model exhibits time non-stationarity while remaining stationary in the frequency domain.

\subsection{Channel Capacity}

The channel capacity, which serves as a crucial metric for evaluating the information transmission capability of the proposed RIS-enabled UAV-to-ship wireless communication system, can be expressed as follows:
\begin{eqnarray}
C = \log_2 \Big( \det\big(\mathbf{I}_{Q} + \frac{\rho_{\text{SNR}}}{P} \mathbf{H}_\text{B} (t, \tau) \mathbf{H}_\text{B}^{\mathrm{H}} (t, \tau) \big) \Big) ,
\end{eqnarray}
Here, $\mathbf{I}_{Q}$ represents the identity matrix of size $Q \times Q$, $\rho_{\text{SNR}}$ represents the signal-to-noise ratio (SNR). This equation reveals that the channel capacity is influenced by various parameters that affect the channel transfer matrix, including the dimensions of the RIS and the partitioning of sub-arrays.

\section{Numerical Results and Discussions}

In this section, we analyze the statistical characteristics of the RIS-enabled UAV-to-vehicle channel propagation by comparing simulation and theoretical results. Here, the modeling accuracy performance is evaluated by the normalized absolute error $\Delta$, which takes the spherical wavefront model as the baseline, written by
\begin{eqnarray}
\Delta = 10 \log_{10} \sum_{q=1}^{Q} \sum_{p=1}^{P} \frac{ | h_{pq}(t,\tau) - h^\text{spherical}_{pq}(t,\tau) | }{ | h^\text{spherical}_{pq}(t,\tau) | } \, .
\end{eqnarray}
Using the sub-array partition framework proposed in this study, we present simulation results for the propagation characteristics of RIS-enabled UAV-to-vehicle channels using the beam-domain and geometry-based methods. Unless otherwise stated, we use the same parameter settings as the channel measurement for our numerical simulations: $P = 30$, $Q = 40$, $K = 1$, $H_0 = 50$ m, $D_0 = 100$ m, $\delta_T = \delta_R = \lambda/2$, $\psi^\text{azi}_T = \pi/3$, $\psi^\text{ver}_T = \pi/4$, $\psi^\text{azi}_R = \pi/3$, $\psi^\text{ver}_R = \pi/4$, $x_\text{RIS} = 50$ m , $y_\text{RIS} = 50$ m, $z_\text{RIS} = 20$ m, $v_T = v_R = 10$ m/s, $\eta^\text{azi}_T = \pi/2$, $\eta^\text{ver}_T = \pi/3$, $\eta_R = \pi/2$, and $t = 1$ s. Furthermore, the RIS parameters are set to $M_x = M_z = 50$ and $d_M = \lambda/2$. In this case, the Fraunhofer distance of the RIS is approximately 150 m, ensuring that both the UAV-RIS subchannel and the RIS-vehicle subchannel fall within the near-field ranges of the~RIS.

\begin{figure}
  \centering
  \includegraphics[width=8.2cm]{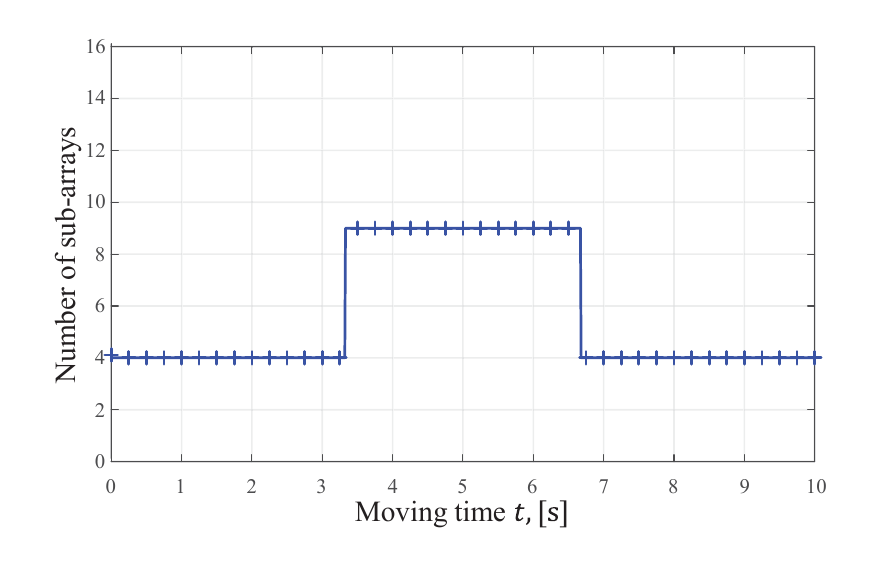}\\
  \caption{Number of sub-arrays in the RIS-enabled UAV-to-vehicle channel model in terms of the different moving time $t$ when $P = 6$ and $Q = 8$.}
  \label{fig_number_of_sub_arrays}
\end{figure}

Fig. \ref{fig_number_of_sub_arrays} shows the evolution of the number of sub-arrays over moving time. It can be found that the number of sub-arrays varies with different moving times, indicating a strong correlation between the proposed sub-array partition analytical framework and the physical characteristics of the RIS. Notably, the number of sub-arrays initially increases with the moving time, reaching a peak of 14, and then decreases after 4 seconds. This changing trend is attributed to the vehicle's motion, initially approaching the RIS and subsequently moving away from it.

\begin{figure}
  \centering
  \includegraphics[width=8.2cm]{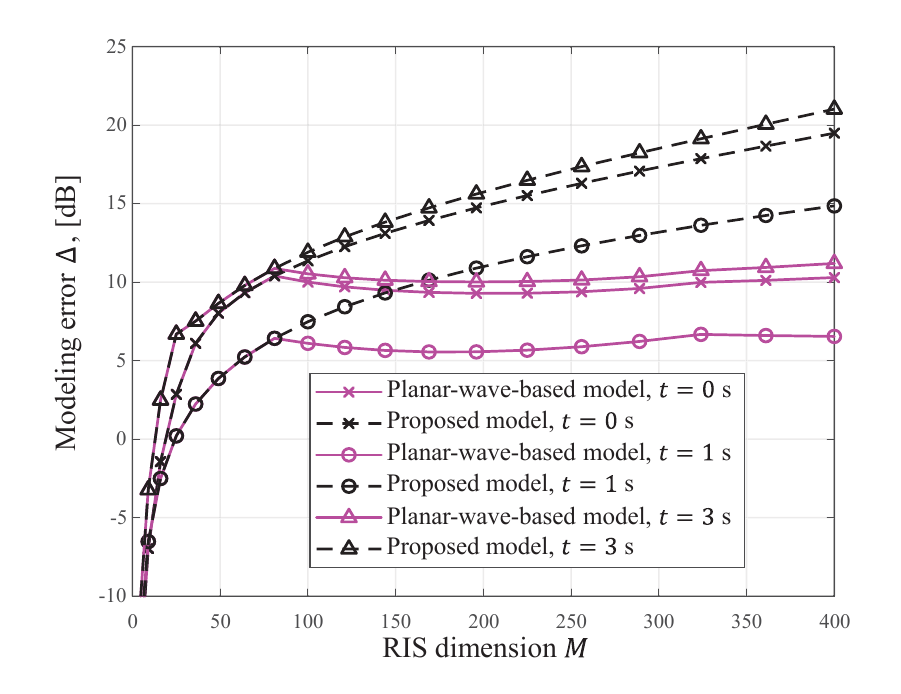}
  \caption{Comparisons between the modeling error performance of the UAV-to-vehicle channel model by using the proposed sub-array partition analytical framework and the planar-wave-based models when $P = 6$, $Q = 8$, $y_\text{RIS} = 30$ m, and $d_M = \lambda / 4$.}
  \label{fig_modeling_accuracy}
\end{figure}

By utilizing (49), Fig. \ref{fig_modeling_accuracy} shows the modeling error performances for various RIS dimensions and different motion time instants. As compared to the conventional planar-wave-based models, the improvements of the proposed channel modeling performance are remarkable, particularly for larger RIS arrays corresponding to near-field ranges. Specifically, when the dimension of the RIS array is limited, resulting in the RIS-enabled UAV-to-vehicle channel model being within the far-field range, the modeling errors based on the proposed framework are the same as those based on the planar-wave-based models. However, with the continually expansion of the RIS dimension, the modeling errors based on the proposed framework will behave better than those based on the planar-wave models. The reason may lie in the fact that when the RIS size is small, which corresponds to the far-field ranges, there no need to operate sub-array partition and therefore the proposed model is substantially equivalent to the planar wavefront model. The above observations highlight the desirable performance of introducing the sub-array partition-based modeling solution for RIS-enabled UAV-to-vehicle systems. Another phenomenon is that the modeling error performances show different behaviors at different time instants, due to the fact that the movement of the UAV and vehicle leads to the changing of the distances between the RIS and UAV/vehicle, and so the Fraunhofer distance is continually varies. The findings in \cite{JTwc} support this observation, providing additional confirmation for the accuracy of the aforementioned~conclusions.

\begin{figure}
  \centering
  \includegraphics[width=8.2cm]{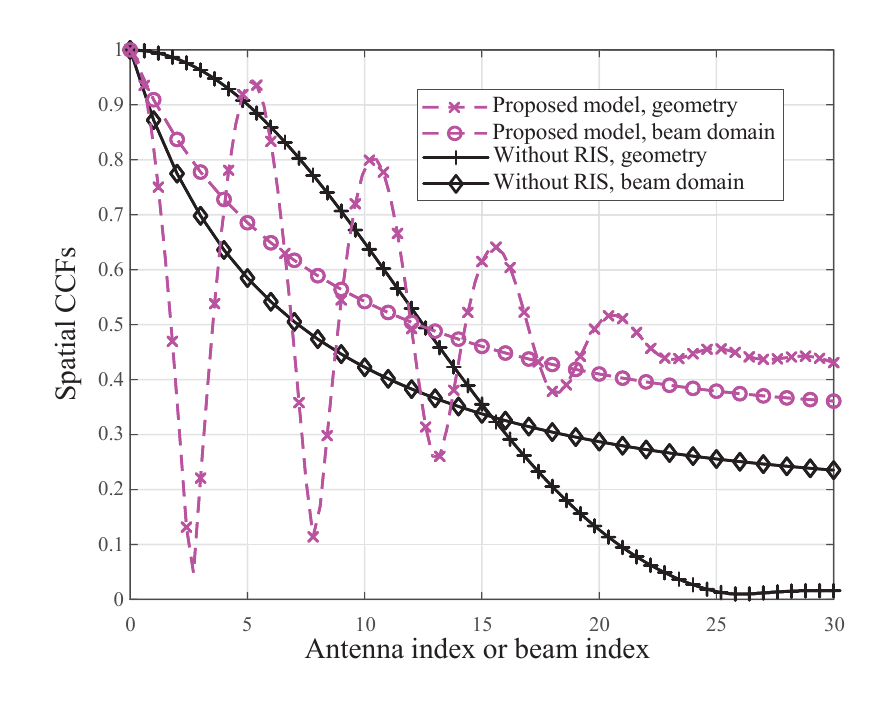}
  \caption{Comparisons between the spatial CCFs of the RIS-enabled UAV-to-vehicle channel model for the cases of consisting and without considering the RIS when $Q = 100$ and $t = 4$ s.}
  \label{fig_CCF_GBSM_BDCM}
\end{figure}

By using Eqs. (43) and (44), the spatial CCFs of the proposed channel model for various modeling methods are depicted in Fig. \ref{fig_CCF_GBSM_BDCM}. It is intriguing to observe that a higher antenna index or beam index results in a reduced spatial correlation, highlighting the non-stationarity of the RIS-enabled UAV-to-vehicle channel in the spatial domain. These simulation observations are in agreement with the measurements from \cite{WTang} and \cite{Payami}; therefore, the accuracy of the derived spatial CCFs in the proposed channel model is confirmed. Due to the sparsity characteristics in UAV-to-vehicle channels, there are noticeable deviations in the spatial correlations between the proposed channel model based on the beam-domain method and the geometry-based method, regardless of whether the waves experience RIS components before reaching the vehicle. Additionally, it is worth noting that the spatial correlations of the proposed channel model increase significantly with the inclusion of RIS components, showcasing the considerable advantages of RIS in enhancing spatial correlations.

\begin{figure}
  \centering
  \includegraphics[width=8.2cm]{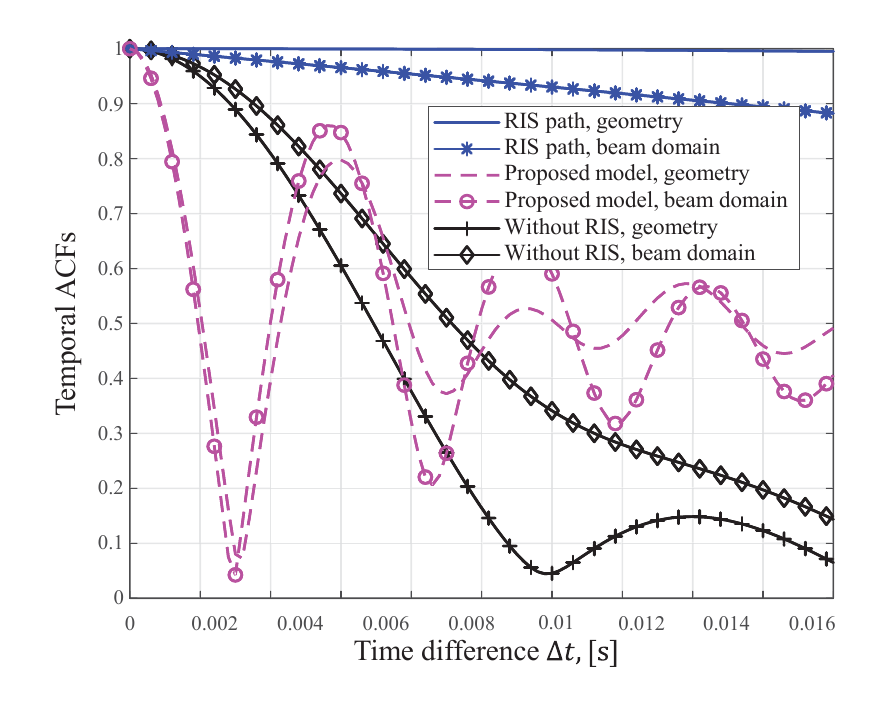}
  \caption{Comparisons between the temporal ACFs of the UAV-to-vehicle channel model for the cases of consisting and without considering the RIS.}
  \label{fig_ACF_GBSM_BDCM}
\end{figure}

Then, we analyze the temporal ACFs of the proposed channel model for various propagation links and modeling methods, as shown in Fig. \ref{fig_ACF_GBSM_BDCM}. Notably, the temporal correlations exhibit varying patterns due to the presence of RIS components in RIS-enabled UAV-to-vehicle channels. This observation aligns with the simulated results in \cite{JTwc}, thus confirming the accuracy of the aforementioned derivation. Additionally, it is evident that the temporal correlations of the proposed channel model decline as the time difference increases, indicating the non-stationarity of the channel in the time domain. Furthermore, there are some deviations between the temporal correlations of the proposed channel model based on the beam-domain method and those based on the geometry-based method, primarily attributed to the sparse features of UAV-to-vehicle wireless channels. Moreover, upon incorporating RIS, the correlations in time domain significantly increase, highlighting the advantage of RIS in enhancing correlations in the time domain.

\begin{figure}
  \centering
  \includegraphics[width=8.2cm]{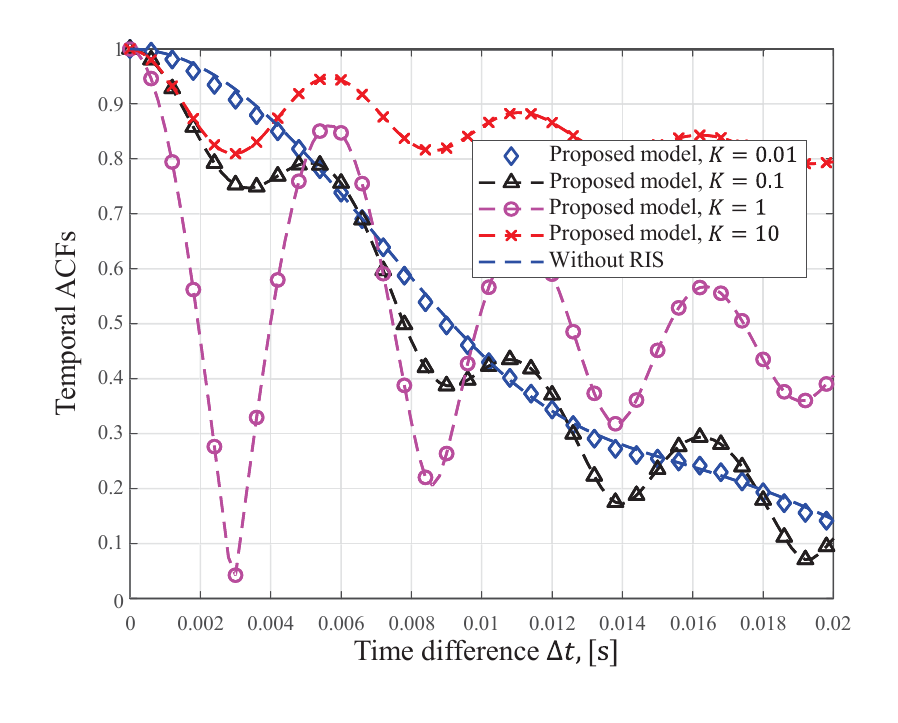}
  \caption{Temporal ACFs of the proposed RIS-enabled UAV-to-vehicle channel model in terms of the different Rician factors $K$.}
  \label{fig_ACF_BDCM_K}
\end{figure}

Existing studies have shown that the percentage of RIS components in the overall links significantly affects channel characteristics \cite{Jhwcl}. In light of this, Fig. \ref{fig_ACF_BDCM_K} illustrates the channel temporal ACFs for RIS-enabled UAV-to-vehicle communications in terms of the Rician factors. It can be observed that as the Rician factor increases from 0.01 to 10, the temporal correlations also increase with the time difference $\Delta t$, which aligns with the findings in \cite{JTcom}. Specifically, when a small Rician factor $K$ is selected, indicating a low proportion of RIS components, the curves of the correlation properties for UAV-to-ground communications exhibit similar behaviors to those without considering RIS. However, it is noteworthy that when $K$ is small, the correlation properties become very large. These observations clearly demonstrate the effectiveness of RIS in enhancing the temporal performance of UAV-to-vehicle communication systems.

\begin{figure}
  \centering
  \includegraphics[width=8.2cm]{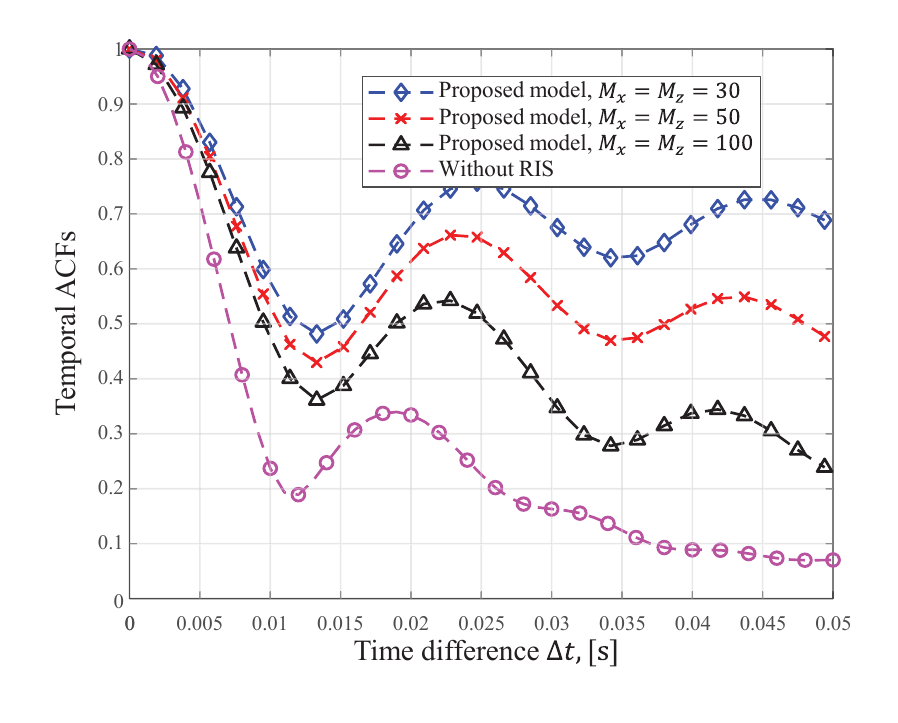}
  \caption{Temporal ACFs of the proposed RIS-enabled UAV-to-vehicle channel model in terms of the different RIS configurations when $t = 4$ s.}
  \label{fig_ACF_BDCM_RIS_size}
\end{figure}

By using Eq. (47), Fig. \ref{fig_ACF_BDCM_RIS_size} illustrates the channel temporal ACFs for RIS-enabled UAV-to-ground communications in terms of the numbers of RIS reflection elements. It can be observed that when RIS is taken into account in UAV-to-vehicle wireless channels, the temporal correlations gradually decrease as the RIS dimension is increased from 30$\times$30 to 100$\times$100. This trend aligns well with the simulated results in \cite{JTwc}; therefore, the accuracy of the derived temporal ACFs in the proposed channel model are confirmed. These findings suggest that the physical features of RIS array have greatly impacts on the propagation characteristics of RIS-enabled UAV-to-vehicle channels. This information can serve as guidance for the efficient assessment of communication systems for UAV-to-vehicle scenarios incorporating RIS technology.

\begin{figure}
  \centering
  \includegraphics[width=8.2cm]{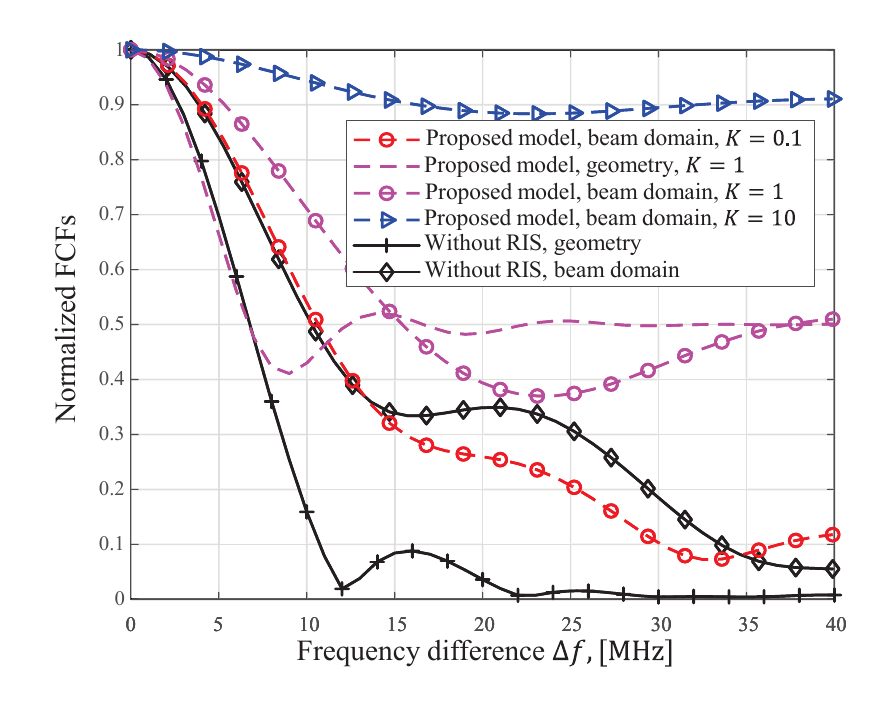}
  \caption{Comparisons between the frequency CFs of the UAV-to-vehicle channel model for the cases of consisting and without considering the RIS.}
  \label{fig_FCF_GBSM_BDCM}
\end{figure}


Fig. \ref{fig_FCF_GBSM_BDCM} shows the channel frequency CFs for RIS-enabled UAV-to-ground communications in terms of the Rician factors. It is evident that a larger value of frequency difference $\Delta f$ results in a lower frequency correlation, indicating the non-stationarity of the channel frequency for RIS-enabled UAV-to-vehicle communications. Specifically, the curves representing the frequency correlations of the propagation components exhibit rapid fluctuations in the initial stage, followed by a tendency towards stabilization as the frequency difference increases. Furthermore, it is noteworthy that the frequency correlations of the RIS-enabled UAV-to-vehicle channel model, based on the beam-domain method, deviate significantly from those based on the geometry-based method, which aligns with the results shown in Fig. \ref{fig_ACF_GBSM_BDCM}. Additionally, the frequency correlations show an increasing trend as the Rician factor $K$ is increased from 0.1 to 10, confirming the findings in Fig. \ref{fig_ACF_BDCM_K}.

\begin{figure}
  \centering
  \includegraphics[width=8.2cm]{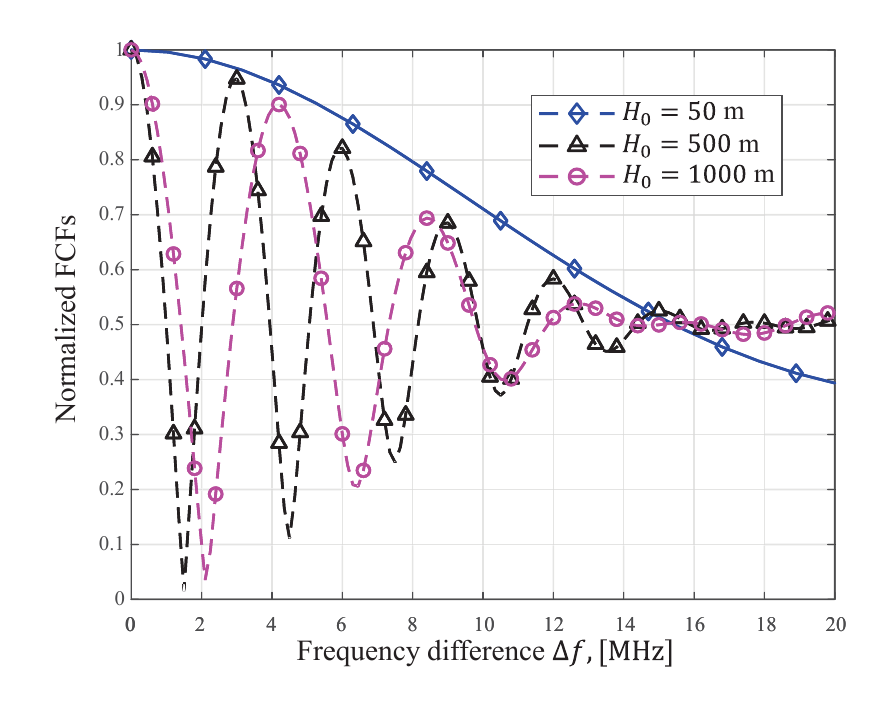}
  \caption{Normalized frequency CFs of the proposed channel model in terms of the different UAV's heights $H_0$.}
  \label{fig_FCF_BDCM_H_0}
\end{figure}

As one of the most crucial aspects in UAV-to-vehicle communication systems, Fig. \ref{fig_FCF_BDCM_H_0} showcases the influence of the UAV's heights on the frequency CFs of RIS-enabled UAV-to-ground wireless channels. It is evident that there are significant variations in the frequency correlations for different UAV height settings. Notably, as the height $H_0$ increases from 50 m to 1000 m, the frequency correlations exhibit a more rapid decrease, which aligns with the findings in the simulations conducted in \cite{Jhwcl}.

\begin{figure}
  \centering
  \includegraphics[width=8.2cm]{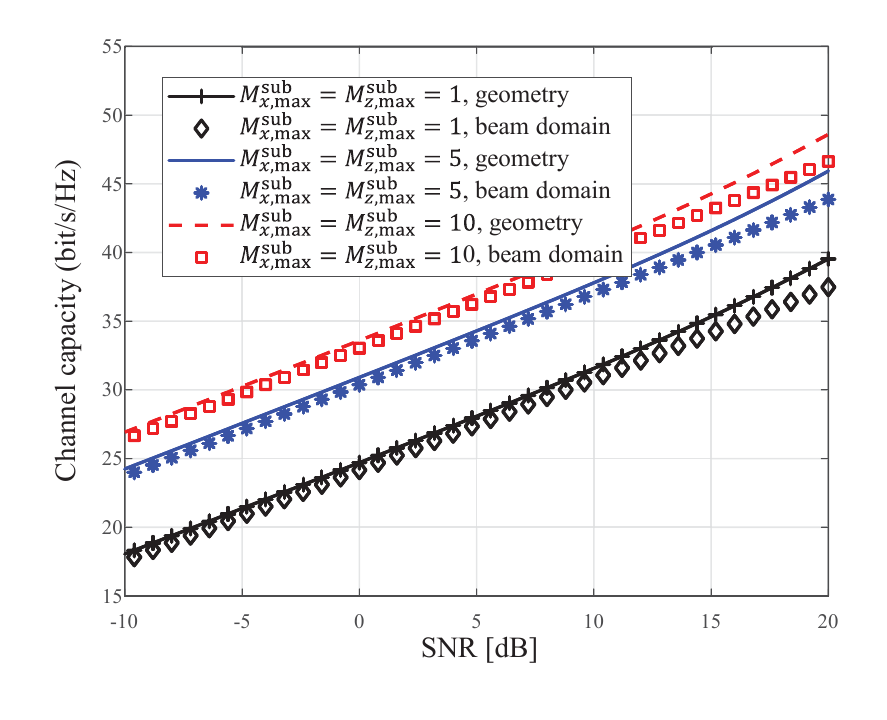}
  \caption{Comparisons between the channel capacities of the RIS-enabled UAV-to-vehicle communication systems by using on the proposed sub-array partition framework in terms of the geometry-based and beam domain methods.}
  \label{fig_channel_capacity_max_sub_array}
\end{figure}

The final set of simulations aims to evaluate the channel capacities of the proposed RIS-enabled UAV-to-vehicle communication systems. By utilizing Eq. (48), Fig. \ref{fig_channel_capacity_max_sub_array} illustrates the channel capacities of the proposed communication system using different modeling methods. It is observed that a higher SNR results in a greater channel capacity for the RIS-enabled UAV-to-vehicle communication system. This finding is consistent with the results reported in \cite{Goldsmith}, which confirms the precision of the calculated channel capacities for the communication system under consideration. Moreover, it is evident that the channel capacities of the RIS-enabled UAV-to-vehicle communication systems, based on the beam-domain method, exhibit some deviations compared to those based on the geometry-based method. These deviations become more pronounced, particularly at high SNR values. Additionally, we observe that the channel capacities increase proportionally as we increase the dimension of the largest sub-array from 1$\times$1 to 100$\times$100, emphasizing the significance of the proposed sub-array partition framework in enhancing system performance.

\begin{figure}
  \centering
  \includegraphics[width=8.2cm]{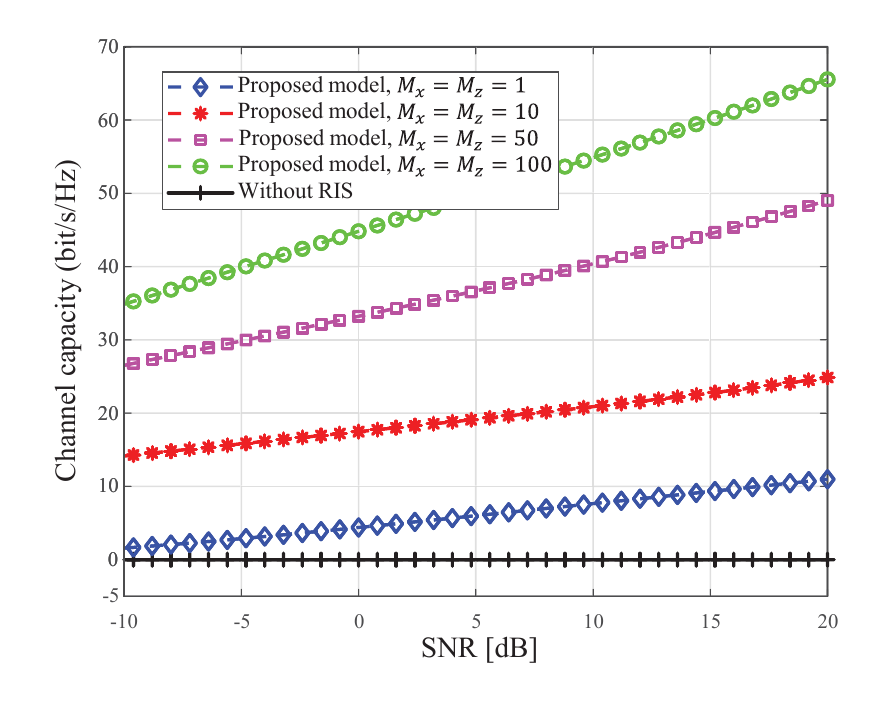}
  \caption{Comparisons of the channel capacities of the UAV-to-vehicle communication systems for the cases of consisting and without considering the RIS.}
  \label{fig_channel_capacity_RIS_size}
\end{figure}

Fig. \ref{fig_channel_capacity_RIS_size} exhibits the channel capacities of the UAV-to-vehicle communication system considering various dimensions of the RIS. Evidently, the presence of RIS components in the proposed channel model results in higher channel capacities. Additionally, the channel capacities of the proposed communication system progressively increase as the dimension of the RIS array grows from 1$\times$1 to 100$\times$100, highlighting the significance of RIS in enhancing channel capacities. It is worth noting that when the RIS dimension exceeds 50$\times$50, the rate at which the channel capacities increase becomes smaller. This phenomenon can be attributed to the saturation of the regulatory capability of RIS. These observations provide valuable insights into the interplay between the physical features of RIS and channel capacities, thereby facilitating the design and assessment of communication systems for UAV-to-vehicle scenarios incorporating RIS technology.

\section{Conclusions}

In this paper, we have presented an analytical framework for sub-array partitioning and subsequently developed a beam domain channel model specifically for near-field communication scenarios in large-scale RIS-enabled UAV-to-vehicle communications. Numerical results demonstrated the effectiveness of our proposed channel model in capturing the sparse features of RIS-enabled UAV-to-vehicle channels, both in near-field and far-field ranges. Our model outperforms conventional methods based on far-field planar-wave-based models, indicating the superiority of our algorithm. Furthermore, we observed that the propagation statistics of RIS-enabled UAV-to-vehicle channels are closely tied to the configurations of the RIS array, specifically the dimensions of the largest sub-array and the total RIS array. These results emphasize the advantages of our framework in balancing complexity and accuracy when investigating the performance of RIS-enabled wireless communication systems, in comparison to existing models.

\end{document}